\documentclass[11pt,letterpaper]{article}
\RequirePackage[T1]{fontenc}
\RequirePackage[utf8]{inputenc}
\RequirePackage{lmodern}
\RequirePackage[letterpaper,margin=1in,includefoot]{geometry}
\RequirePackage{amsmath,amssymb,amsfonts,mathrsfs}
\RequirePackage{graphicx,xcolor,textcomp,array,booktabs,siunitx,tabularx,chemformula}
\RequirePackage{subcaption}\RequirePackage[hyphens]{url}
\RequirePackage[numbers,sort&compress,merge,round]{natbib}
\RequirePackage{jabbrv}
\RequirePackage{microtype}
\RequirePackage{etoolbox}
\RequirePackage{titlesec}
\RequirePackage{fancyhdr}
\RequirePackage{hyperref}
\hypersetup{hidelinks}

\AtBeginEnvironment{thebibliography}{\interlinepenalty=10000}
\titleformat{\section}{\normalfont\fontsize{14}{17}\selectfont\bfseries}{\thesection}{1em}{}
\titleformat{\subsection}{\normalfont\fontsize{12}{14}\selectfont\bfseries}{\thesubsection}{1em}{}
\titleformat{\paragraph}[runin]{\normalfont\normalsize\bfseries}{\theparagraph}{1em}{}
\fancypagestyle{plain}{\fancyhf{}\fancyfoot[C]{\small\thepage}}
\usepackage{authblk}

\usepackage[utf8]{inputenc}
\usepackage{amsmath,amssymb,amsfonts}
\usepackage{mathrsfs}
\usepackage{graphicx}
\usepackage{xcolor}
\usepackage{textcomp}
\usepackage{array}
\usepackage{booktabs}
\usepackage{siunitx}
\usepackage{tabularx}
\usepackage{chemformula}
\usepackage{subcaption}
\usepackage{url}
\usepackage[T1]{fontenc} 
\DeclareSIUnit{\year}{yr}
\newcommand{\safeincludegraphics}[2][]{%
  \IfFileExists{#2}{%
    \includegraphics[#1]{#2}%
  }{%
    \fbox{\parbox[c][0.18\textheight][c]{0.9\linewidth}{\centering
      \textbf{Figure asset unavailable in this workspace}\\[0.5em]
      \footnotesize\path{#2}}}%
  }%
}
\newcommand{\lnUwrap}[2]{\shortstack[l]{$\ln U(#1,$\\$#2)$}}
\newcommand{\Uwrap}[2]{\shortstack[l]{$U(#1,$\\$#2)$}}
\newcolumntype{L}[1]{>{\raggedright\arraybackslash}p{#1}}
\newcolumntype{Y}{>{\raggedright\arraybackslash}X}
\newcolumntype{C}[1]{>{\centering\arraybackslash}p{#1}}
\newcommand{\valstack}[1]{\shortstack[l]{#1}}

\makeatletter
\renewcommand{\maketitle}{%
  \begin{center}
  {\fontsize{14}{17}\selectfont\bfseries\@title\par}\vspace{1em}
  {\@author\par}
  \end{center}
  {\small\noindent\preprintcorrespondence\par}\vspace{1em}
}
\makeatother
\newcommand{\leadauthor}[1]{}
\newcommand{\dates}[1]{}
\newcommand{\significancestatement}[1]{\long\def\preprintsignificance{#1}}
\newcommand{\authorcontributions}[1]{\long\def\preprintcontributions{#1}}
\newcommand{\authordeclaration}[1]{\def\preprintdeclaration{#1}}
\newcommand{\correspondingauthor}[1]{\def\preprintcorrespondence{#1}}
\newcommand{\keywords}[1]{\def\preprintkeywords{#1}}
\newcommand{\matmethods}[1]{\long\def\preprintmethods{#1}}
\newcommand{\showmatmethods}{\section*{Materials and Methods}\preprintmethods}
\newcommand{\dataavail}[1]{\section*{Data, Materials, and Software Availability}#1}
\newcommand{\acknow}[1]{\long\def\preprintacknow{#1}}
\newcommand{\showacknow}{\section*{Acknowledgments}\preprintacknow}

\RequirePackage{longtable}
\RequirePackage{pdflscape}
\RequirePackage{needspace}
\RequirePackage{caption}
\newcommand{\makesititle}{%
  \thispagestyle{plain}
  \vspace*{0.3in}
  {\noindent\fontsize{14}{18}\selectfont\bfseries Supporting Information for\par}
  \bigskip
  {\noindent\fontsize{14}{18}\selectfont\bfseries\SIembeddedtitle\par}
  \bigskip
  {\noindent\normalsize\SIembeddedauthors\par}
  \bigskip
  {\noindent\normalsize\SIcorrespondence\par}
  \bigskip
  \noindent\textbf{This supplement includes:}
  \begin{itemize}
    \item Supporting text
    \item Figs. S1 to S6
    \item Tables S1 to S3
    \item SI References
  \end{itemize}
  \clearpage
}
\newenvironment{sifigure}{%
  \clearpage\begingroup
  \renewcommand{\caption}[1]{%
    \par\medskip\needspace{3\baselineskip}%
    \begingroup\small\parindent=0pt\leftskip=0pt\rightskip=0pt\parfillskip=0pt plus 1fil
    \refstepcounter{figure}\textbf{Fig.\ \thefigure.}\space ##1\par
    \endgroup}
}{\par\endgroup}
\newcommand{\SIembeddedtitle}{Fermentation heterotrophy can reproduce Enceladus's plume \ch{H2} and \ch{CH4} under limited serpentinization}
\newcommand{\SIembeddedauthors}{Sam Nasreldine, Antonin Affholder, Laura Molares Moncayo, Régis Ferrière, and François Guyot}
\newcommand{\SIcorrespondence}{Corresponding authors: Sam Nasreldine and Régis Ferrière.\\E-mail: snasreldine@arizona.edu, regisf@arizona.edu}

\begin{document}
\pdfbookmark[0]{Main manuscript}{main-manuscript}

\title{Fermentation heterotrophy can reproduce Enceladus's plume \ch{H2} and \ch{CH4} under limited serpentinization}

\author[a,1]{Sam Nasreldine}
\author[b]{Antonin Affholder}
\author[c,d,i]{Laura Molares Moncayo}
\author[e,f,g,1]{Régis Ferrière}
\author[h]{François Guyot}

\affil[a]{Program in Applied Mathematics, University of Arizona, Tucson, AZ, USA}
\affil[b]{Centre for the Origin and Prevalence of Life, Department of Environmental Systems Science, ETH Zurich, Zurich, Switzerland}
\affil[c]{School of Geography, Queen Mary University of London, London, UK}
\affil[d]{Natural History Museum, London, UK}
\affil[e]{International Research Laboratory for Interdisciplinary Global Environmental Studies (iGLOBES), CNRS, ENS--PSL University, and University of Arizona, Tucson, AZ, USA and Paris, France}
\affil[f]{Department of Ecology and Evolutionary Biology, University of Arizona, Tucson, AZ, USA}
\affil[g]{Institut de Biologie de l'École Normale Supérieure (IBENS), École Normale Supérieure, Université PSL, Paris, France}
\affil[h]{Institut de Minéralogie, de Physique des Matériaux et de Cosmochimie, Muséum National d'Histoire Naturelle, Paris, France}
\affil[i]{Aix Marseille Université, Université de Toulon, CNRS, IRD, MIO, Marseille, France}

\leadauthor{Nasreldine}

\significancestatement{Enceladus ejects \ch{H2} and \ch{CH4} from a subsurface ocean, raising the possibility of hydrothermal ecosystems. Previous models have shown that these gases' concentrations in Enceladus's plume could plausibly be explained by methanogens fueled by serpentinization-derived hydrogen, but sustained global \ch{H2} production may be limited. We show that an alternative microbial metabolism, fermentation of abiotically-produced organic substrates, can reproduce Cassini fluxes under limited serpentinization. Successful cases are concentrated in acetate-routing, fermentation-supported communities rather than purely hydrogenotrophic ecosystems, motivating future searches for acetate, ammonium, organic intermediates, and methane isotopic signatures.}

\authorcontributions{S.N., A.A., L.M.M., R.F., and F.G. conceived the study. S.N. and A.A. developed the model and code, with contributions from L.M.M. and F.G. S.N. conducted the simulations, analyzed the results, and wrote the first draft with input from all coauthors. All authors reviewed and finalized the manuscript. R.F. provided funding.}
\authordeclaration{The authors declare no competing interests.}
\correspondingauthor{\textsuperscript{1}To whom correspondence should be addressed. E-mail: snasreldine@arizona.edu, regisf@arizona.edu}

\keywords{Enceladus $|$ astrobiology $|$ fermentation $|$ methanogenesis $|$ Bayesian inference}

\dates{}

\maketitle
\begin{abstract} Saturn's icy moon Enceladus is a prime target in the search for life beyond Earth. Abundant \ch{H2} and \ch{CH4} in Enceladus's south-polar plume have been shown to be consistent with hydrogenotrophic methanogens feeding on products of long-lived serpentinization. However, thermo-permeability models make sustained core-wide \ch{H2} production by serpentinization uncertain. We therefore explore whether an alternative metabolism based on fermentation-driven heterotrophy could reproduce Cassini \ch{H2} and \ch{CH4} escape fluxes under limited serpentinization. We develop a coupled ecosystem model of fermentation and methanogenesis in a hydrothermal-ocean-plume framework, using glycine as a proxy for amino-acid-like fermentable organics. With effectively no abiotic \ch{H2} and \ch{CH4} input, purely heterotrophic communities with fermenters and methanogens can reproduce both Cassini-derived flux intervals given sufficient hydrothermal glycine-equivalent supply, while the estimated organic stock could sustain the required \ch{H2} production on gigayear timescales if fully available. We then test a scenario with input ranges for fermentable organic matter, \ch{H2}, and \ch{CH4} informed by asteroid-sample measurements and geochemical estimates of limited serpentinization. Simulations within both Cassini-derived flux intervals are concentrated in communities with acetoclastic methanogens; Bayesian model comparison favors the same communities. This holds even allowing for greater abiotic \ch{H2} and \ch{CH4} inputs that reduce the fermentable-organic supply required to reproduce the plume. We therefore identify fermentation heterotrophy as a testable alternative to purely hydrogenotrophic, serpentinization-dominated biotic interpretations and motivate future plume searches for co-exported fermentation products, including acetate and ammonium, together with measurements of methane carbon isotopes. \end{abstract}
\noindent\textbf{Keywords:} \preprintkeywords

\section*{Significance}
\preprintsignificance

\section*{Introduction}

NASA's Cassini mission revealed a global liquid water ocean within Enceladus's subsurface and performed \textit{in situ} chemical analyses of material derived from it via ejection in the south polar plume.
This first ever analysis of material from an extraterrestrial liquid water environment revealed  abundant dihydrogen (\ch{H2}) and methane (\ch{CH4}) \cite{Waite2017}.
Together with the expectation set by models for hydrothermal flow through Enceladus's rocky core powered by tidal dissipative heating \cite{Choblet2017}, and empirical arguments \cite{hsu2015ongoing}, the presence of these volatiles has been interpreted as evidence for hydrothermal alteration of the rocky core resembling Earth's serpentinization process.
These findings also served as a basis for formulating a specific scenario for Enceladus's habitability.
Earth's hydrothermal vents harbor methanogens that obtain energy for growth and carbon fixation through hydrogenotrophic methanogenesis, the reduction of carbon dioxide (\ch{CO2}) by serpentinization-derived \ch{H2} to methane.
Therefore, the scenario of Earth-like methanogens as candidate organisms for Enceladus's interior came into focus as a central hypothesis for Enceladus's habitability \cite{Waite2017,taubner2018biological,Higgins2021,Affholder2021}.

However, sustained intense serpentinization required to balance loss of \ch{H2} in the plume \cite{Hansen2011,Waite2017,Hansen2020} over the expected age of tidal heating in Enceladus's core \cite{Nimmo2023} clashes with the relatively short timescales expected for exhausting the budget of weatherable minerals within Enceladus's core \cite{Daval2022,Zandanel2021}.
This challenges the ability of serpentinization to sustain the plume's \ch{H2} flux and support Enceladus's habitability over gigayear timescales.
Could there be scenarios of interior processes at Enceladus, other than serpentinization, to explain the observed present day release of \ch{H2} via the plume and support a consistent hypothesis for the habitability of its subsurface?

Accretionary material incorporated into outer Solar system objects is considered to be analogous in composition to carbonaceous chondrites and comets, and therefore expected to be rich in organic matter \cite{Lichtenberg2021}, and the moment of inertia of Saturn's moons is consistent with an organic component in the core \cite{Nri2020, Reynard2023}.
This has led others, in the case of Titan, to propose that some of the accretionary organic matter could be released in Titan's presumed ocean and might support metabolisms sourcing carbon and energy from organic matter (heterotrophic metabolisms) \cite{Miller2026,Affholder2025}.
Similarly, Enceladus's core is likely to have an organic component which may have undergone or still undergoes dissolution into the liquid water ocean. 
This suggests that considering a scenario of Enceladus's ocean habitability being supported by heterotrophy, i.e. metabolisms that source carbon and energy from organic substrates, is warranted, in addition to (chemo)autotrophic scenarios such as hydrogenotrophic methanogenesis (where carbon is sourced from inorganic \ch{CO2}).

Anoxic conditions are expected to dominate in Enceladus's ocean in the absence of known abiotic processes that would generate abundant dioxygen (\ch{O2}), with the absence of light penetration in Enceladus's ocean further rendering a hypothetical photosynthetic source of \ch{O2} implausible. Under these conditions, anaerobic heterotrophic respiration requires an external terminal electron acceptor to oxidize organic matter (e.g., elemental sulfur or iron (III)). Radiolytic processes may deliver some oxidizing power to Enceladus's ocean \cite{Ray2021}. Our budget calculation indicates that their capacity to oxidize methane, averaged over 4.5 Gyr, is small relative to the adopted plume methane flux (SI Appendix, Supplementary Results S2). On this basis, we omit methanotrophy from the ecosystem model.
In contrast, fermentation conserves energy without requiring an external terminal electron acceptor, instead maintaining redox balance through organic substrates and their metabolic intermediates.
Fermentation therefore offers a mechanism by which Enceladus's putative organic inventory could support heterotrophic habitability.

Terrestrial hydrothermal systems provide a useful analogue for assessing plausible heterotrophic metabolisms on Enceladus. Alongside diverse autotrophic communities, hydrothermal vent ecosystems harbour metabolically versatile anaerobic heterotrophs capable of exploiting a wide range of biotically and abiotically produced organic substrates through fermentation \cite{Zeng2021,Zhou2022}.
In Earth's hydrothermal ecosystems, methanogens and anaerobic fermenters often form mutually beneficial metabolic partnerships. Fermenters supply substrates for methane production, while methanogens consume these products, keeping their concentrations low enough for fermentation to remain energetically favorable \cite{Jackson2002,HoldenSistu2023}.

\begin{sloppypar}
Fermentation reactions produce dihydrogen \ch{H2}, together with acetate/acetic acid (\ch{CH3COO^-/CH3COOH}), ammonium/ammonia (\ch{NH4+ / NH3}) and dissolved inorganic carbon (\ch{CO2_{(aq)} + HCO3^- + CO3^{2-}}), while methanogenesis, a secondary process carried out by other microbes, can use either hydrogen (hydrogenotrophic methanogenesis) or acetate (acetoclastic methanogenesis) to produce methane \cite{Schink1997}. For example, fermenters may supply acetate to acetoclastic methanogens in the hydrothermally altered sediments of the Guaymas Basin \cite{Su2023}. The coupled activity of fermenters and methanogens therefore provides a mechanism seemingly compatible with the presence of hydrogen and methane in Enceladus's plume which are hard to reconcile with known abiotic processes.
\end{sloppypar}

Quantitatively evaluating this hypothesis requires mechanistic modeling of specific metabolisms supported by empirical knowledge of the microorganisms that perform them \cite{Affholder2021,Higgins2021,Affholder2025}.
Not all organic molecules are known to be substrates for fermentation, and it is likely that the particular redox chemistry of fermentation strongly constrains the range of molecules that can be fermented and metabolised.
Conservatively, we therefore focus on a specific fermentation metabolism, carried out by well-characterised microorganisms occupying environmental niches analogous to those proposed for Enceladus's ocean. We must also consider a metabolism which conserves energy and carbon from an organic molecule, or a class of metabolic molecules, expected to be present at appreciable levels in Enceladus's ocean, either from dissolution of accretionary material or from \textit{de novo} abiotic synthesis.

Amino acids in particular can serve as energy sources, commonly fermented in pairs, with one amino acid oxidized and another reduced, and ATP synthesized by substrate-level phosphorylation following deamination \cite{deVladar2012,Schonheit2016}. Fermentation of a single amino acid is also possible. Evidence for glycine metabolism in hydrothermal environments includes both cultured deep-sea isolates, including \textit{Clostridium tepidiprofundi} \cite{Slobodkina2008,Poehlein2016}, and metagenome-assembled genomes recovered from alkaline serpentinite-hosted hydrothermal systems encoding a glycine reduction pathway inferred to consume \ch{H2} \cite{Nobu2023}. We therefore propose \ch{H2}-producing glycine (\ch{NH2-CH2-COOH}) fermentation as a model heterotrophic metabolism for Enceladus, motivated by evidence for microbial glycine use in hydrothermal environments analogous to those proposed for Enceladus's ocean and the expected availability of glycine within the accretionary organic inventory of outer Solar System bodies \cite{Affholder2025}.

This approach is consistent with the analysis of samples returned from asteroids Bennu and Ryugu, which builds on decades of studies of organic-rich meteorites.
Soluble organics found in these asteroidal samples include amino acids, amines, carboxylic acids, sugars, and nucleobases, and may result from aqueous alteration of insoluble organic matter (IOM) in their parent bodies or low-temperature photochemistry in interstellar or protosolar ice grains \cite{GlavinBennu2025,Potiszil2023,Parker2023,Naraoka2023Sci,Furukawa2025}.
These findings are relevant to the emergence of life on Earth and suggest that readily fermentable organics could have been incorporated into the organic-bearing material in Enceladus's core.
Amino acids, foremost among which glycine, may be the most abundant class of molecules known to be fermentable in carbonaceous chondrites and their parent asteroids \cite{Glavin2010,GlavinBennu2025}. Enceladus may also have accreted a cometary component \cite{Nri2020}; glycine has been detected in cometary material \cite{Altwegg2016}.

We estimate a possible initial glycine inventory for Enceladus from measurements in primitive Solar System material and assumptions about its core composition (Methods, Calculating feasibility of fermentation).
The escape rate of \ch{H2} in Enceladus's plume is approximately $(1.2\text{--}7.4)\times10^9$\,\si{\mol\per\year} \cite{Waite2017,Hansen2011,Hansen2020}.
Glycine fermentation follows the stoichiometric equation \ref{eq:glycine_fermentation} \cite{Thauer1977,Affholder2025}:
\begin{equation}
\label{eq:glycine_fermentation}
\begin{aligned}
\ch{NH2CH2COOH + 2 H2O}
&\longrightarrow \ch{1/2 CH3COOH + NH3} \\
&\quad \ch{+ CO2_{(aq)} + H2O + H2}. %
\end{aligned}
\end{equation}
Fermentation of the estimated initial glycine stock could sustain the observed plume hydrogen escape rate on gigayear timescales, assuming that the entire stock were available for fermentation.
This makes the case for the heterotrophic habitability of Enceladus's ocean particularly interesting to consider, whereas it has so far received only limited attention.

The lifetime of glycine in Enceladus's ocean may, however, be relatively short if hydrothermal circulation is ancient and vigorous, owing to abiotic degradation which occurs otherwise slowly at low, bulk-ocean, temperatures \cite{Truong2019}.
Abiotic degradation of organic matter in Enceladus's core could also produce the volatiles detected in the plume, as there is no fundamental distinction between thermodynamically favorable reactions used by life for energy conservation, and those occurring spontaneously.
Biological interpretations of these volatiles must therefore be evaluated against the range and uncertainty of Enceladus's abiotic baseline \cite{Higgins2026}. Any proposed biological scenario should accordingly generate quantitative, testable predictions that can be compared with plume observations. The hypothesis of a fermentation-supported microbial community at Enceladus's seafloor raises such questions beyond the stoichiometric and inventory arguments presented above.
Is the volatile composition of Enceladus's plume an expected output from Earth-like fermentative communities?
How does community composition affect the expected community outputs, and if so, which community composition would best explain methane and hydrogen escape rates in the plume?
Can ongoing dissolution of accretionary glycine supply a microbial community inhabiting seafloor hydrothermal vents with sufficient organics to match rates of hydrogen and methane production implied by plume escape rates?

\begin{figure}[p]
  \centering
  \begin{subfigure}[c]{0.55\textwidth}
    \centering
    \safeincludegraphics[width=\linewidth]{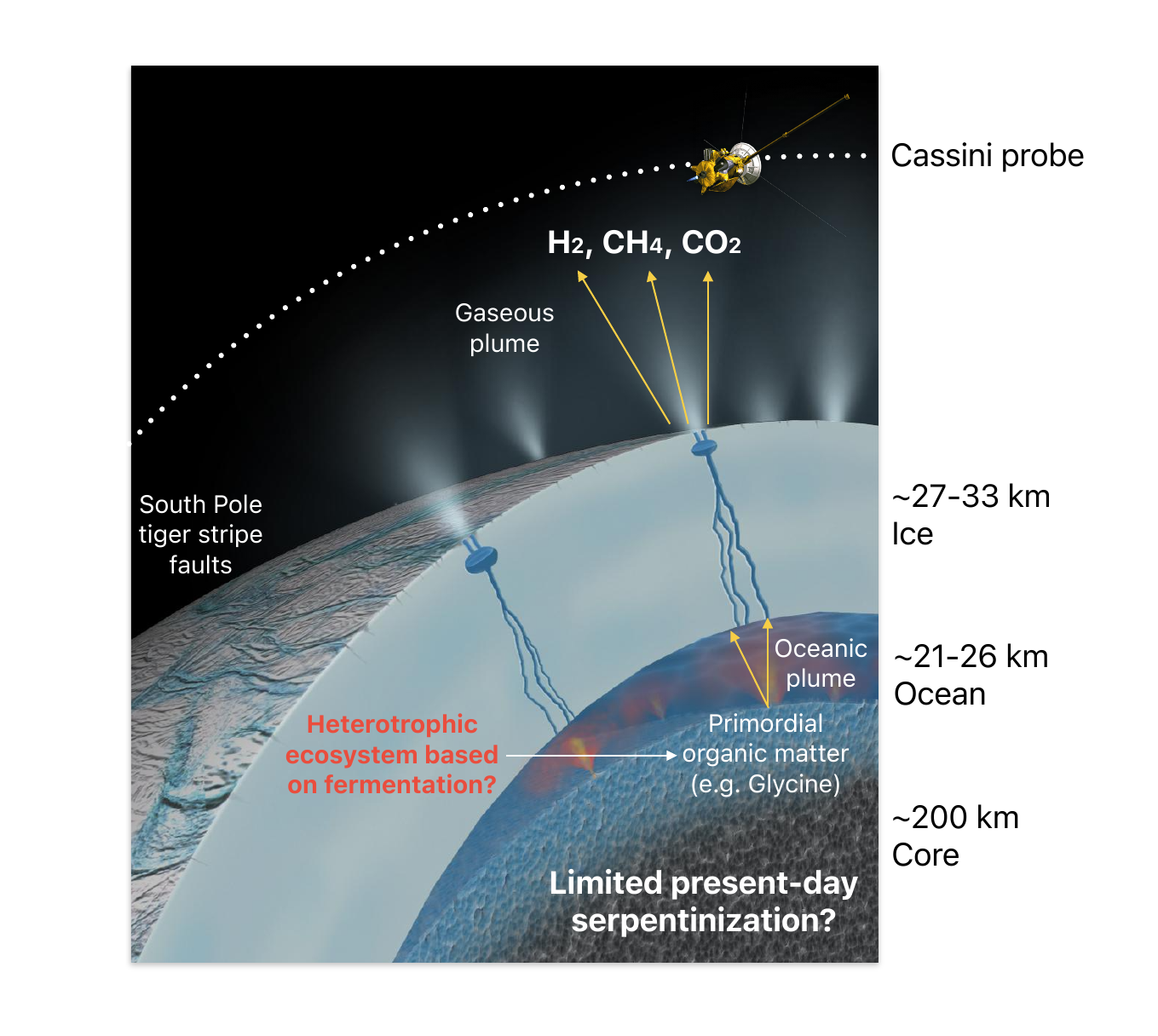}
    \caption{}
  \end{subfigure}
  \hfill
  \begin{subfigure}[c]{0.44\textwidth}
    \centering
    \safeincludegraphics[width=\linewidth]{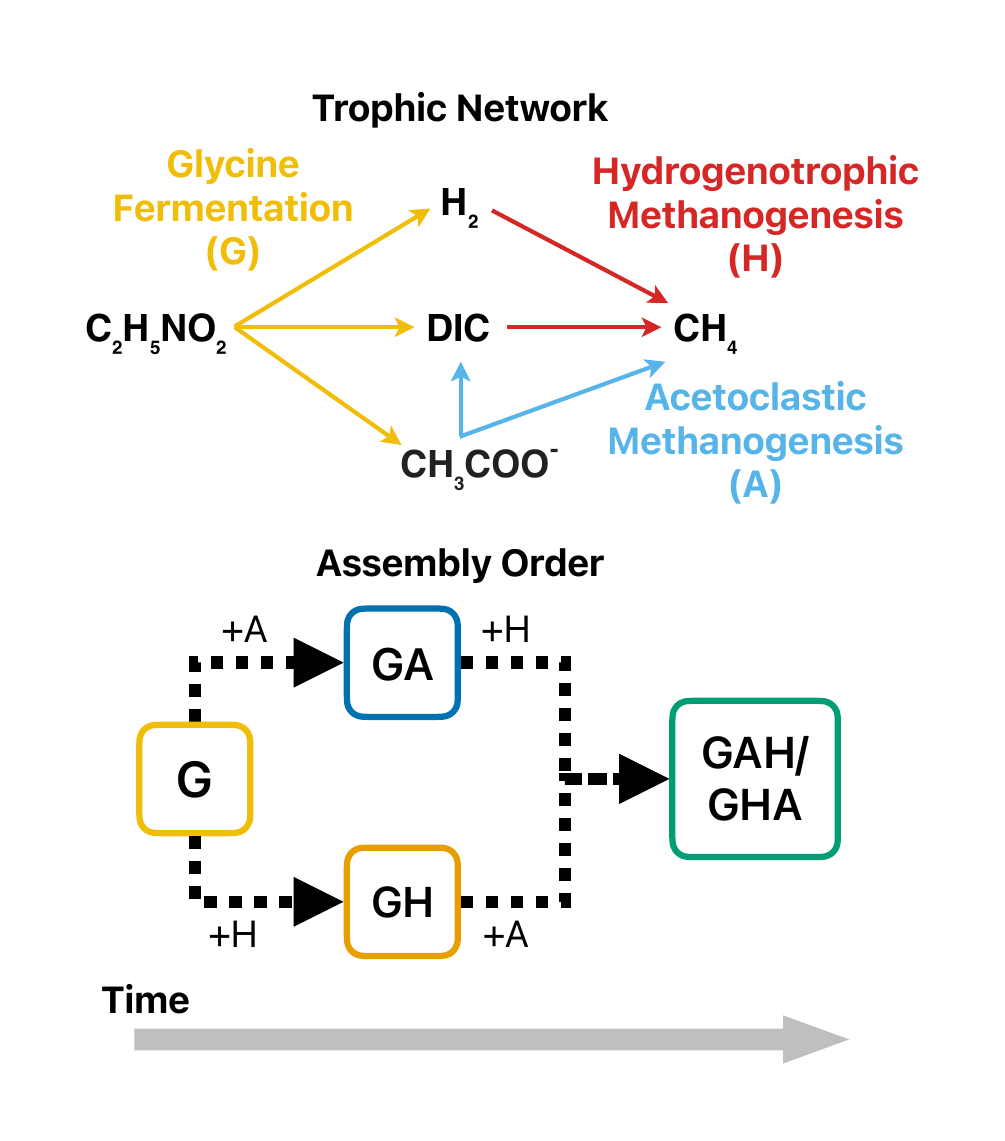}
    \caption{}
  \end{subfigure}
  \caption{%
    \textbf{A heterotrophic ecosystem enabled by fermentation on Enceladus.}
    \textbf{(a)} \textbf{Conceptual cross-section of Enceladus's interior and heterotrophic ecosystem hypothesis.} We use the interior dimensions of Park et al. \cite{Park2024} for the schematic: a silicate core of $\sim$200 km radius, an ocean of $\sim$21--26 km mean thickness, and an ice shell of $\sim$27--33 km mean thickness. Limited present-day serpentinization in the core and a putative ecosystem produce \ch{H2}, \ch{CH4}, \ch{CO2}, and aqueous products, which ascend with hydrothermal fluids through South Polar tiger-stripe faults into the ice and erupt as plumes. Primordial organic matter (e.g., glycine) accreted in the core could support a heterotrophic ecosystem based on fermentation; here glycine serves as a proxy for fermentable organics with similar end products. Yellow arrows indicate transport of dissolved gases and aqueous products into the oceanic plume. Cassini's Ion and Neutral Mass Spectrometer (INMS) sampled gas in the plume ejected from Enceladus's south polar region \cite{Waite2017}. The biomass in the drawing illustrates the proposed ecosystem. Simulated biomass remains in the mixing layer; the INMS comparison uses only \(\Phi_{H_2}\) and \(\Phi_{CH_4}\). Background image \cite{NASA_JPL_PIA19058} and Cassini spacecraft illustration courtesy of NASA/JPL-Caltech.
    \textbf{(b, top)} \textbf{Trophic network among modeled guilds}: colored arrows show the substrate/product links by which the three modeled metabolisms interact. Glycine fermentation supplies \ch{H2}, acetate, and dissolved inorganic carbon (DIC) pools used by methanogens; hydrogenotrophic methanogenesis consumes \ch{H2} and DIC to produce \ch{CH4}, whereas acetoclastic methanogenesis consumes acetate and produces \ch{CH4} and DIC. Additional products, including \ch{NH4+}/dissolved inorganic nitrogen (DIN) from glycine fermentation, are included in the model stoichiometry but omitted from the schematic for clarity.
    \textbf{(b, bottom)} \textbf{Modeled community assembly pathways}: staged simulations introduce glycine fermenters first, followed by hydrogenotrophic methanogens, acetoclastic methanogens, or both, yielding GH, GA, and mixed GHA/GAH consortia.}
  \label{fig:ecosystem_schema}
\end{figure}

To investigate these questions, we build upon an existing bioenergetic model for microbial growth \cite{sauterey2020,Affholder2021,Affholder2025} to simulate the ecology and ecosystem function of a minimal fermentative community that processes \ch{H2} and \ch{CH4} (Fig.~\ref{fig:ecosystem_schema}).
We couple this model to an idealized simulation of the mixing between hydrothermal fluid and ocean water around a seafloor hydrothermal vent \cite{Affholder2021} to simulate consumption of organics and production of volatiles at the scale of the seafloor hydrothermal environment. 
We then map different hypotheses for abiotic processes within Enceladus's core, ongoing weak or strong serpentinization, efficiency of the dissolution of glycine into hydrothermal fluid, ocean composition and pH, to simulation parameters, i.e., temperature and chemical composition of the hydrothermal fluid and composition of the ocean.
Then, by exploring how different metabolic assemblages are permitted, not only by the abiotic description of the hydrothermal environment, but also by biotically constructed habitability in the form of hydrogen and acetate produced by fermenters and fueling the metabolism of methanogens, we investigate the ecological factors that constrain the hypothesis of a seafloor fermentative community as a candidate explanation for the composition of Enceladus's plume.

\section*{Results}
We organized the simulations into three principal cases, each isolating a distinct geochemical question via their associated priors sampled.
We represent fermentable organic matter with glycine and calculate its fermentation products using Eq.~\ref{eq:glycine_fermentation}. Glycine-equivalent supply refers to the amount of this modeled substrate supplied per year, expressed in moles of glycine. Each mole contains two moles of carbon. The Minimal case tests how much of this supply is needed to reproduce both Cassini-derived flux intervals with negligible abiotic inputs. Abiotic \ch{H2}, \ch{CH4}, and acetate concentrations in the ocean and hydrothermal fluids are fixed at a numerical floor of \SI{1e-12}{\mol\per\kilogram}, so glycine fermentation supplies essentially all modeled \ch{H2} and subsequent \ch{CH4} production depends entirely on fermentation products.
The Standard case tests whether the same community types can reproduce the plume under conservative priors that assume limited sustained global serpentinization (SI Appendix, Table S1) in addition to constrained release of glycine at the seafloor. The Serpentinization-Enhanced case tests how increased abiotic inputs of hydrogen and methane change the required fermentable organic supply and the relative support for each community type. Additional sensitivity analyses isolate the effects of high \ch{H2}, glycine supply alone, acetate supply alone, and combined glycine--acetate supply alone. Hereafter, G denotes glycine fermenters, H denotes hydrogenotrophic methanogens, and A denotes acetoclastic methanogens (Fig.~\ref{fig:ecosystem_schema}). H-only simulations provide a comparison for methane production without fermentation. We group simulations by the guilds tested: H, G+H (GH), G+A (GA), or all three guilds. We record separately which guilds remain at the end of each simulation. For the three-guild simulations, GHA denotes adding hydrogenotrophic and then acetoclastic methanogens after the glycine fermenters, and GAH denotes the reverse methanogen order. The model-choice analysis includes only simulations in which each guild could grow when introduced and the model reached equilibrium after each addition (SI Appendix, Methods S3--S4).

\begin{figure}[p]
\centering

\includegraphics[width=\textwidth]{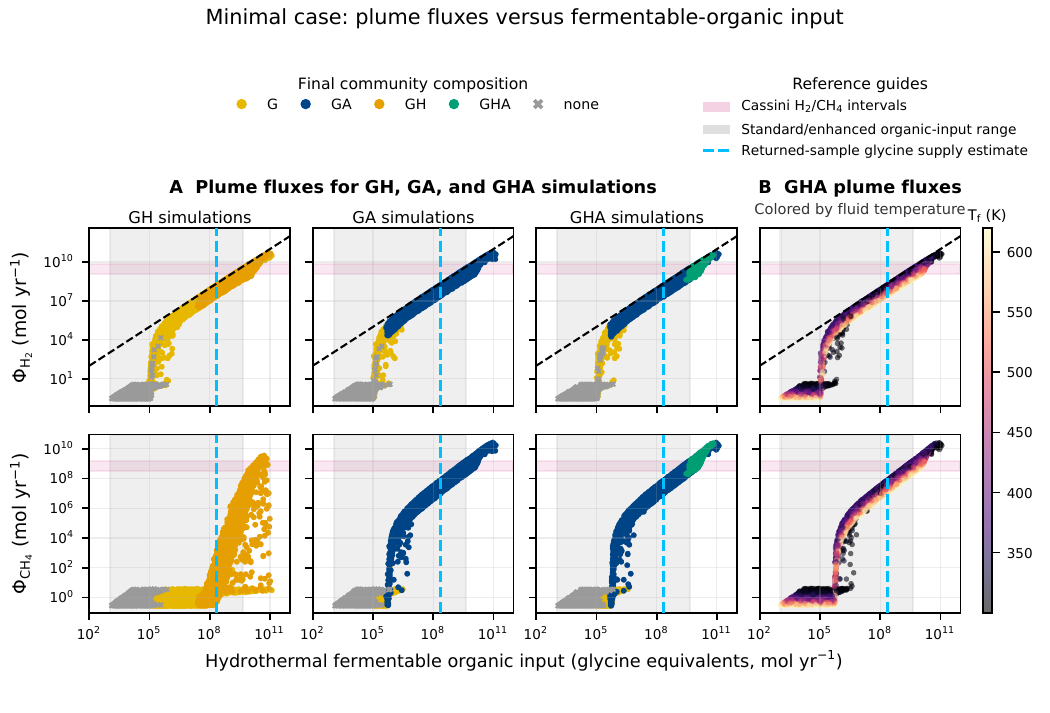}

\caption{\textbf{Minimal-case plume fluxes versus fermentable-organic input.} (A) Simulated annual \ch{H2} and \ch{CH4} plume fluxes plotted against hydrothermal fermentable-organic input for GH, GA, and GHA simulations. Points are colored by the final community composition: G (yellow), GA (blue), GH (orange), GHA (green), or no surviving guild (gray). (B) The GHA simulations from panel A colored by hydrothermal-fluid temperature, \(T_f\). The dashed black lines give the limit of complete conversion of supplied glycine into fermentation product \ch{H2} according to the stoichiometry in Eq.~\ref{eq:glycine_fermentation}. Pink horizontal bands show the Cassini-derived intervals for hydrogen and methane escape rate. The gray vertical bands show the fermentable-organic input range calculated from the Standard and Serpentinization-Enhanced glycine concentration priors, and the dashed cyan lines mark the upper annual glycine supply calculated using the glycine concentration estimated from Bennu samples and the maximum modeled hydrothermal flow (SI Appendix, Methods S4--S5).}
\label{fig:t1_efficiency}
\end{figure}

We first examined the Minimal case using glycine stoichiometry, which isolates how efficiently a fermentable donor pool can be translated into plume \ch{H2} and \ch{CH4} (Fig.~\ref{fig:t1_efficiency}). More glycine generally produces more \ch{H2} and \ch{CH4} in GH, GA, and GHA simulations (Fig.~\ref{fig:t1_efficiency}A).
Simulations that predict methane escape rates within the Cassini window mostly correspond to guild assemblages GA and GHA, both of which include acetoclastic methanogens. In contrast, simulations with the GH assemblage (fermentation and hydrogenotrophic methanogenesis) yield much less methane.
At a fixed glycine supply, changing hydrothermal-fluid temperature changes GHA gas production efficiency by approximately one order of magnitude (Fig.~\ref{fig:t1_efficiency}B).

\paragraph*{Substantial glycine supply is required to produce gigamoles of \ch{H2} and \ch{CH4} annually}

GH simulations reproduce both Cassini-derived flux intervals much less often than GA or GHA simulations.
With glycine fermentation supplying essentially all \ch{H2}, the simulations that reproduce both Cassini-derived flux intervals require hydrothermal glycine-equivalent concentrations in the upper part of the Minimal prior (GA/GHA: \(1.7\times10^{-3}\)--\(1.0\times10^{-1}\)\,\si{\mol\per\kilogram}; GH: \(1.7\times10^{-2}\)--\(1.0\times10^{-1}\)\,\si{\mol\per\kilogram}). Our back-of-the-envelope calculation suggests that the estimated initial glycine stock could supply the observed hydrogen escape rate on gigayear timescales, assuming that the entire stock is available for fermentation and all hydrogen produced escapes through the plume (Methods, Calculating feasibility of fermentation).

However, supplying the required concentrations using only glycine at the abundance measured in Bennu samples remains a challenge. Even assuming complete release into 0.3 kg of water per kilogram of material, we estimate an upper dissolved concentration of \(1.5\times10^{-4}\)\,\si{\mol\per\kilogram} (SI Appendix, Methods S5), below the concentrations required by the simulations that reproduce both gases.

\paragraph*{Communities with acetoclastic methanogens better reproduce both plume fluxes than hydrogenotrophic alternatives}

Under Standard priors, 91 GA simulations and 91 GHA simulations out of 30,000 samples satisfy both Cassini-derived flux constraints (0.30\% for each); no GH or H simulations do. Hydrogenotrophic methanogens consume dissolved \ch{H2} to produce \ch{CH4}. For a given hydrogen supply, producing more methane therefore leaves less hydrogen for export, making it harder for GH and H simulations to reproduce both Cassini-derived flux intervals. Under Standard priors, neither methanogen introduction order produced a final community containing hydrogenotrophs (SI Appendix, Supplementary Results S4). The simulations therefore support acetoclastic methanogenesis in two ways: communities containing acetoclastic methanogens reproduce both observed gas fluxes, and these methanogens remain viable under conditions where hydrogenotrophs do not persist. Acetoclastic methanogens can instead convert acetate to \ch{CH4} without consuming \ch{H2}. Leaving this \ch{H2} available for export helps these simulations reach both observed intervals more often.

\paragraph*{Fermentation enables additional vent-scale habitat for methanogens}

We compared simulated growth with and without fermentation for each methanogenic guild at each position in the modeled vent. We recorded a habitat gain when fermentation made growth of methanogens possible at any position where the guild could not grow under the abiotic background. The SI Appendix gives the areas gained and lost and the net change, calculated using the area represented by each position. In the Minimal test case, fermentation newly enables either hydrogenotrophic or acetoclastic methanogenesis somewhere in the vent in 19,907 of 29,970 completed simulations initiated by glycine fermentation (66.4\%). Acetoclastic methanogenesis is enabled in 19,907 simulations (66.4\%), hydrogenotrophic methanogenesis in 10,699 simulations (35.7\%), and both guilds gain newly viable habitat somewhere in the same simulation in 10,699 simulations (35.7\%). Under Standard priors, enablement of either methanogen population occurs in 2,804 of 29,994 simulations (9.3\%; hydrogenotrophs 7.3\%, acetoclastic methanogens 2.6\%, both guilds 0.58\%). Under Serpentinization-Enhanced priors, it occurs in 2,023 of 29,942 simulations (6.8\%; hydrogenotrophs 1.9\%, acetoclastic methanogens 5.2\%, both guilds 0.32\%). These results show that fermentation can make additional vent locations permissive for methanogens even when methanogenesis is not solely dependent on fermentation products.

\paragraph*{What plume constraints select within the viable simulations}

The simulations from the Standard set, with guild assemblages GA/GHA that yield methane and hydrogen outputs consistent with Cassini-derived intervals require large glycine supply and moderate or low hydrogen supply (SI Appendix, Fig.~S1).
In the Serpentinization-Enhanced case, where higher concentrations of abiotic \ch{H2} are allowed, the glycine required for simulations to match with hydrogen and methane intervals is relaxed in comparison to the Standard set (SI Appendix, Fig. S1).
Within each case, the subsets of simulations with guild assemblages GA and GHA have nearly identical flux and input medians, suggesting that selecting simulations matching the plume composition favors the scenario of fermentation-supported acetotrophy, independently of its association with hydrogenotrophy.

For each parameter sample, we simulated both methanogen introduction orders, with glycine fermenters introduced first. This tested whether the absence of hydrogenotrophs from the final community depended on which methanogen was introduced first. Under Standard priors, hydrogenotrophs that persisted alongside glycine fermenters were absent after acetoclastic methanogens were tested in the next stage. Reversing the order of methanogen introduction gave the same result: hydrogenotrophs were absent from the final community in all 29,994 simulations. The area where acetoclastic methanogens could grow remained nearly unchanged. Neither order therefore produced a final community containing all three guilds under Standard priors (SI Appendix, Supplementary Results S4).

\begin{figure}[p]
  \centering
  \safeincludegraphics[width=\textwidth]{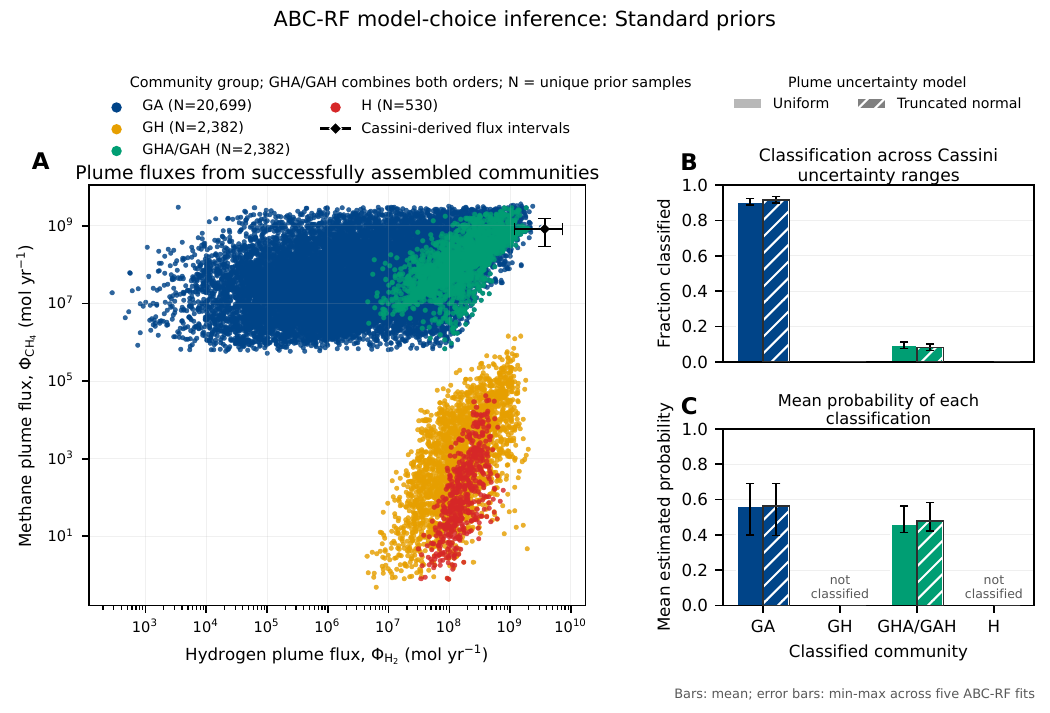}
  \caption{\textbf{ABC--RF model choice under Standard priors.} We tested 30,000 parameter samples under these priors. (A) Plume \ch{H2} and \ch{CH4} fluxes from simulations in which each guild could grow when introduced and the model reached equilibrium after each addition. The legend gives the number of unique prior samples: GA, 20,699; GH, 2,382; GHA/GAH, 2,382; and H, 530. The pooled GHA/GAH group combines the two methanogen introduction orders after glycine fermenters. If only one order met the conditions for a prior sample, it received that sample's full weight; if both did, each received half. Under Standard priors, all 2,382 prior samples in this group contributed only the GHA order. The labels describe the guilds introduced and do not necessarily describe the final community. The black diamond shows the fluxes calculated from the arithmetic midpoints of the adopted water-ejection and gas-abundance ranges. The error bars span the adopted flux interval for each gas. The midpoint calculations use a linear scale, so the diamond lies off-centre within the intervals on the logarithmic axes. All simulations meeting the stated conditions were retained, and the four groups received equal total weight. (B) Fraction of flux pairs sampled from the Cassini uncertainty ranges that ABC--RF assigns to each community group. (C) Estimated probability that a classification is correct, averaged over the flux pairs assigned to each group. These probabilities are estimated using the regression forest described in SI Appendix, Methods S7. In both panels, bars show means across five fits with different random seeds, and error bars span their minimum and maximum. Uniform and truncated-normal sampling represent alternative assumptions about uncertainty in the water-ejection rate and gas abundances.}
  \label{fig:abc_standard}
\end{figure}

\begin{figure}[p]
  \centering
  \safeincludegraphics[width=\textwidth]{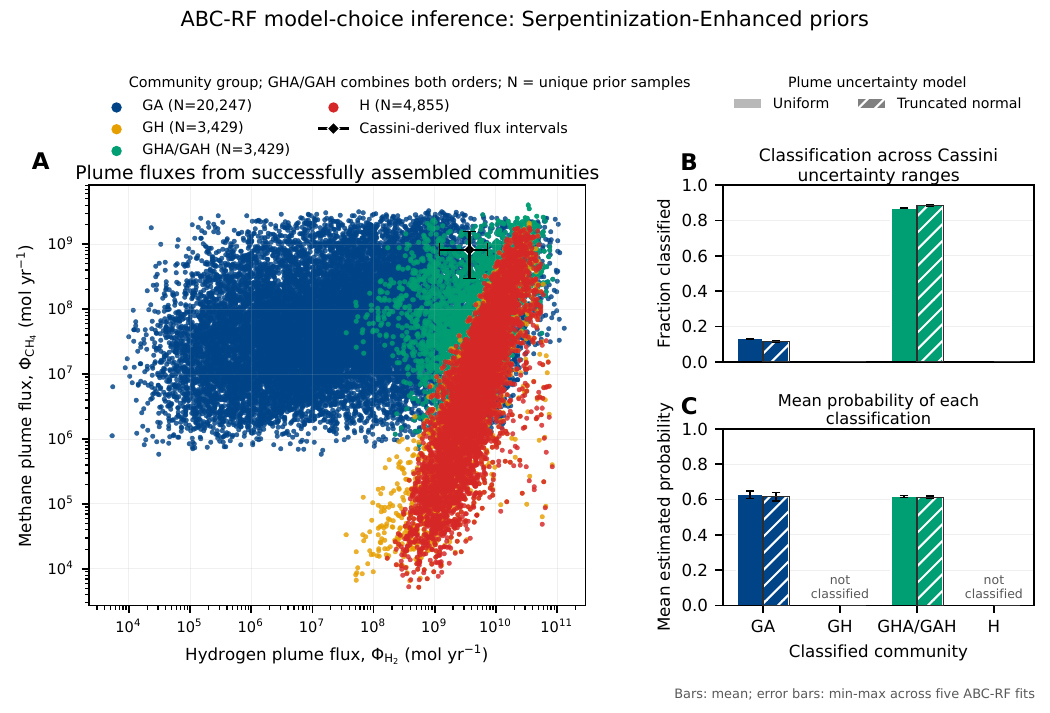}
  \caption{\textbf{ABC--RF model choice under Serpentinization-Enhanced priors.} These priors allow broader abiotic \ch{H2} and higher abiotic \ch{CH4} inputs than the Standard case. We tested 30,000 parameter samples under these priors. (A) Plume \ch{H2} and \ch{CH4} fluxes from simulations in which each guild could grow when introduced and the model reached equilibrium after each addition. The legend gives the number of unique prior samples: GA, 20,247; GH, 3,429; GHA/GAH, 3,429; and H, 4,855. The pooled GHA/GAH group contains 5,000 simulations from 3,429 unique prior samples. Only the GHA order contributed for 1,858 prior samples. Both orders contributed for 1,571 samples, with half the sample's weight assigned to each order. A label records which guilds were introduced; those guilds need not all remain in the final community. The black diamond shows the fluxes calculated from the arithmetic midpoints of the adopted water-ejection and gas-abundance ranges. The error bars span the adopted flux interval for each gas. The midpoint calculations use a linear scale, so the diamond lies off-centre within the intervals on the logarithmic axes. We included every simulation meeting the stated conditions and gave the four groups equal total weight. (B) Fraction of flux pairs sampled from the Cassini uncertainty ranges that ABC--RF assigns to each community group. (C) Estimated probability that a classification is correct, averaged over the flux pairs assigned to each group. These probabilities are estimated using the regression forest described in SI Appendix, Methods S7. In both panels, bars show means across five fits with different random seeds, and error bars span their minimum and maximum. Uniform and truncated-normal sampling represent alternative assumptions about uncertainty in the water-ejection rate and gas abundances.}
  \label{fig:abc_serp}
\end{figure}

We compared the ability of simulations with guild assemblages H, GH, GA, and GHA/GAH to explain the combined flux of \ch{H2} and \ch{CH4} in Enceladus's plume using approximate Bayesian computation with random forests (ABC--RF; Figs.~\ref{fig:abc_standard} and \ref{fig:abc_serp}).
We included simulations with at least one modeled vent location where each guild had non-negative net growth when introduced and the system reached equilibrium after each addition (SI Appendix, Methods S3--S4). The H, GH, GA, and GHA/GAH groups describe which guilds were viable when introduced rather than which persisted at equilibrium. An introduced guild can alter substrate and product concentrations during assembly even if it later dies off. The pooled group combines the two orders for adding the methanogens after glycine fermenters: hydrogenotrophic methanogens before acetoclastic methanogens (GHA), or the reverse (GAH). Each parameter sample received a total weight of 1 within the pooled GHA/GAH group. When both introduction orders met the growth and equilibrium criteria, each simulation received a weight of 0.5; when only one did, it received the full weight. This retained both orders while giving each parameter sample equal representation. We then rescaled the weights so that H, GH, GA, and GHA/GAH each contributed 25\% of the total training weight. Our aim was to assess how the plume measurements distinguish these four alternatives after applying the growth and equilibrium criteria. Equal group weights give each alternative the same initial support, preventing differences in the number of qualifying simulations from determining their prior probabilities.
We generated pairs of \ch{H2} and \ch{CH4} plume fluxes across the intervals constrained by Cassini, using the uncertainty models described in SI Appendix, Methods S7. For each flux pair, ABC--RF predicted the most likely community group among the four biological alternatives, under the adopted priors. Under Standard priors, which assume limited abiotic \ch{H2} and \ch{CH4} inputs, ABC--RF assigned 90.6--91.7\% of these flux pairs to GA and 8.3--9.4\% to GHA/GAH; none were assigned to GH or H.
Under priors of the Serpentinization-Enhanced case, 11.5--13.1\% were classified as GA and 86.9--88.5\% as GHA/GAH; again, none were classified as GH or H. For each flux pair, ABC--RF also estimated the probability that its community assignment was correct, given the four biological alternatives considered. Averaged over flux pairs assigned to each group and then across five fits, these probabilities were 0.56 for GA and 0.46--0.48 for GHA/GAH under Standard priors. Under Serpentinization-Enhanced priors, they were 0.62--0.63 for GA and 0.61--0.62 for GHA/GAH. The ranges reflect uniform versus truncated-normal sampling of the Cassini uncertainties.

For each prior sample, we compared the two orders in which hydrogenotrophic and acetoclastic methanogens were added. In all three cases, the same guilds remained at the end in at least 99.9\% of comparisons. Hydrogen flux differed by less than about 0.23\% between the two introduction orders. Methane flux differed by more than about 2.3\% between the two introduction orders in four Minimal comparisons, but in none under Standard or Serpentinization-Enhanced priors. The largest difference was about 10\% (SI Appendix, Fig. S6).

In the Minimal case, fermentation produces both observed gas fluxes without abiotic \ch{H2}, \ch{CH4}, or acetate input when the glycine-equivalent supply is in the upper part of the explored range. We also tested communities supplied with glycine and acetate, using hydrothermal concentration ranges based on abundances measured in asteroid samples and assumptions about their dissolution into water. Abiotic \ch{H2} and \ch{CH4} input concentrations were set to effectively zero (\(10^{-12}\ \mathrm{mol\,kg^{-1}}\)) in both hydrothermal fluid and ocean water (SI Appendix, Methods S5 and Table S2). None of these simulations reproduced both Cassini-derived flux intervals (SI Appendix, Fig. S2). With acetate alone, \ch{CH4} can reach its interval, but \ch{H2} remains below its observed range (SI Appendix, Fig. S4). GA and GHA simulations reach both intervals in the high-\ch{H2} sensitivity (SI Appendix, Fig. S3).

\section*{Discussion}

Our modeling brings interest for a relatively underappreciated scenario of Enceladus's habitability, where biomass production is supported by heterotrophs consuming primordial organic matter.
Our simulations suggest that physicochemical conditions in Enceladus's interior may favorably support the growth of known organisms that consume organic matter anaerobically and under reducing conditions, satisfying a key aspect of habitability \cite{cockell2016habitability}.

A back of the envelope calculation suggests that the primordial glycine inventory could sustain the observed plume hydrogen flux through fermentation on the timescale of over a billion years  (Methods, \emph{Calculating feasibility of fermentation}) \cite{Waite2017,Hansen2011,Hansen2020}. These timescales are comparable to estimates for the duration of vigorous tidal heating at Enceladus \cite{Nimmo2023}, making fermentation worth investigating as a potentially long-lived process. Limited access to the stock of fermentable organics would shorten this lifetime.

More broadly, our modeling examines how habitability is shaped by the abiotic environment and by microbial activity.
In our simulations, fermentation can support methanogenesis when the abiotic supply of hydrogen (\ch{H2}) and acetate is scarce, but the supply of glycine or similar molecules is not. This occurs even in the Minimal case, in which abiotic hydrogen and acetate inputs are effectively zero.
Under Standard priors, hydrogenotrophs were absent from the final community regardless of methanogen introduction order, consistent with exclusion by acetoclastic methanogens (SI Appendix, Supplementary Results S4). Hydrogenotroph loss also occurs under Serpentinization-Enhanced priors, although both methanogenic guilds can persist together in either introduction order.
This tackles a distinctly original aspect of habitability, assessing how individual metabolic guilds can be assumed to be present within a community, consistently with the presence of other metabolic guilds. The broader implication is that habitability can itself be modified by life: organisms change the chemical conditions that support or limit other metabolisms. The gases a biosphere produces consequently depend on the community that develops as organisms transform their environment. By calculating substrate production and consumption together with microbial growth and community assembly, our approach makes these ecological interactions part of the biosignature prediction itself. Predicting signals from such biospheres therefore requires understanding both the resources supplied by the planetary environment and the communities that can persist while transforming them.

This significantly affects how abiotic factors determine the ability of methanogenic communities to produce both \ch{H2} and methane (\ch{CH4}) at rates matching the rate at which these volatiles are removed from the ocean via plume eruption.
Indeed, the main outcome of this modeling investigation is that a fermentation-supported community can produce \ch{CH4} and \ch{H2} at rates comparable to their eruption rates, without requiring methane production to be driven mainly by long-lived, globally sustained serpentinization. 
In simulations of communities containing glycine fermenters and acetoclastic methanogens (GA), and optionally hydrogenotrophic ones (GHA), for which \ch{H2} and \ch{CH4} production match plume escape rates, methanogenic substrates are provided by fermentation. This reduces the requirement for abiotic supply of those substrates, and therefore may allow more \ch{H2} to be left unconsumed while simultaneously allowing methane to be produced in abundance.
The loss of hydrogenotroph persistence in these Standard-prior simulations could also help maintain high \ch{H2} export even when hydrogenotrophy is included among the modeled metabolisms.

For our modeled communities, the concentration of organic matter available in hydrothermal fluids is also a key factor. Glycine abundances measured in returned asteroid samples, combined with assumptions about release from rock into water, provide conditional bounds on this supply (SI Appendix, Methods S5). Under Minimal priors, simulations that reproduce both gas fluxes require glycine concentrations above the upper bound inferred under those assumptions (Fig.~2). Their plausibility therefore depends on whether a greater fermentable supply could be available.

The larger upper bound explored in our simulations allows glycine to represent the combined supply of glycine and other similarly fermentable compounds. These could include organics released from the core or synthesized abiotically \cite{Craddock2026}, with a possible additional contribution from organic matter produced by an earlier autotrophic methanogenic population \cite{Affholder2022}. Whether these sources could sustain the larger inputs remains an important uncertainty.

With glycine inputs limited to the bounds inferred from asteroid samples, an additional abiotic \ch{H2} supply may be needed to reproduce both observed gas fluxes. More abundant reactive minerals and exposed alteration surfaces could have supported greater \ch{H2} production through serpentinization in the past \cite{Daval2022,Zandanel2021}. Serpentinization may still contribute to the abiotic \ch{H2} supply today, albeit at lower rates than in the past. This possibility motivated the abiotic \ch{H2} range used in our Standard priors. Continued serpentinization would depend on the availability of reactive minerals \cite{Daval2022,Zandanel2021}. Tidal fracturing could also generate abiotic \ch{H2} when newly exposed mineral surfaces react with water. Sustaining this source over geological timescales would require broken mineral bonds to reform between fracturing events \cite{Oberg2026}.

In GA and GHA simulations, fermentation supplies acetate for methane production.
We can test this link further by measuring the compounds exported with the gases.
Simulations within both Cassini-derived intervals also export acetate, dissolved inorganic nitrogen (ammonia and ammonium), and dissolved inorganic carbon, sometimes with unconsumed glycine (SI Appendix, Fig. S1). Measurements of these compounds' abundances relative to methane could test whether plume chemistry is consistent with the consumption and production predicted for fermentation-supported communities. Interpreting these measurements would require accounting for changes during transport from the vent environment to the plume and for detection limits. Methane carbon-isotope measurements could help assess which pathways produced the methane. Their interpretation requires accounting for the effects of pathway and substrate on isotopic fractionation, as demonstrated in cultures of thermophilic and acetoclastic methanogens \cite{Valentine2004,Londry2008}. Hydrogen availability can also affect fractionation during hydrogenotrophic methanogenesis \cite{Okumura2016}. Assessing a biological contribution would also require comparison with the carbon-isotope signatures expected from abiotic processes, which remain uncertain for Enceladus \cite{Higgins2026}.

The choice of simulated observables remains a central approximation of the comparison with Cassini data. As in Affholder et al. (2021) \cite{Affholder2021}, the comparison is made at the level of outgoing \ch{H2} and \ch{CH4} fluxes rather than concentrations inferred for the plume source material. We prescribe ocean and hydrothermal-fluid compositions and calculate their mixing near the seafloor. We then estimate annual gas fluxes from the resulting dissolved concentrations. Comparing these fluxes with plume escape assumes that the amounts of \ch{H2} and \ch{CH4} in the ocean remain approximately constant, with hydrothermal export balanced by escape through the plume \cite{Affholder2022}. We do not simulate the subsequent transport of dissolved material, grains, or cells through a stratified ocean and individual tiger-stripe fractures, or biological reactions during that transport. During transport from the hydrothermal source to the plume, these reservoirs may be fractionated relative to one another \cite{Ames2025}. Gas exsolution can change the relative abundances of gases entering the plume \cite{Fifer2022}, while water-vapor condensation can increase the proportions of other gases relative to water \cite{Fifer2022}. Sorting of chemically distinct ice grains \cite{Postberg2011} and collisions between grains and fracture walls \cite{Schmidt2008} also affect the particulate material sampled by a spacecraft. Biological products may also degrade during transport. Whether they remain measurable depends on the balance between biological production and abiotic removal, as discussed by Neveu et al. \cite{Neveu2018}.

Our predictions for the composition of material leaving the seafloor also depend on how hydrothermal flow and chemical reactions near the vents are modeled. We base core flow on the permeable-core model of Choblet et al. \cite{Choblet2017}. Rock alteration enters only through the inlet composition \(C_i^0\). Within the mixing layer, amino-acid losses are limited to biological consumption and transport out with the fluid. Thermal degradation, which can occur in hot hydrothermal fluids and solutions relevant to ocean worlds, would provide an additional sink \cite{Bada1995,CoxSeward2007,Truong2019}. Including it could reduce the substrate available for fermentation, particularly in hot fluids with long residence times. For numerical convenience, oxidants are also kept outside the state variables of Eq.~\ref{eq:main_ode_system}; averaging the optimistic oxidant budget of Ray et al. \cite{Ray2021} over 4.5 Gyr gives a methane-oxidation capacity equivalent to 0.5--2.6\% of the adopted plume methane escape rate (SI Appendix, Supplementary Results S2).

A major biological simplification of this work is the use of a single fermentative guild \(G\). In Eq.~\ref{eq:main_ode_system}, \(G\) denotes the glycine-fermentation metabolism of Eq.~\ref{eq:glycine_fermentation} \cite{Thauer1977}.  Anaerobic degradation of amino-acid-rich organic matter in terrestrial systems can involve additional fermentative reactions, including Stickland-type oxidation--reduction pairs \cite{deVladar2012,Schonheit2016}, as well as hydrolysis of polymeric organic matter before fermentable monomers become available and tighter or more diverse syntrophic coupling to \ch{H2}-consuming partners than represented here \cite{Schink1997,Stams2009}. These processes could alter substrate availability and the production of \ch{H2} and acetate, thereby affecting downstream methane production.

We use the single glycine fermentation reaction (Eq.~\ref{eq:glycine_fermentation}) to specify the carbon, nitrogen, and hydrogen balance in the ODE system. Extending the calculation to the other amino acids, sugars, and small organic molecules reported in Bennu samples \cite{GlavinBennu2025,Furukawa2025} would require stoichiometries, kinetic laws, and thermodynamic constraints for the additional reactions. Separate guilds would also be needed where substrate use requires different physiology. Glycine provides a stoichiometric proxy for that broader fermentable pool. Its use is motivated by amino-acid organics in carbonaceous material, asteroids, and comets \cite{GlavinBennu2025,Glavin2010,Altwegg2016}, and follows earlier models of anaerobic heterotrophy on ocean worlds \cite{Affholder2025}. 

Carbon incorporated into biomass is removed from the dissolved chemistry, and its subsequent return to organic or inorganic reservoirs lies outside the model. We use this treatment to compare \ch{H2} and \ch{CH4} plume fluxes among communities.
The consideration of heterotrophy in models of hypothetical Enceladean biospheres should motivate extending modeling approaches to track carbon and nutrient cycling.
Existing biomass estimates are sensitive to habitat scale and growth-limitation assumptions. Nutrient shortages can restrict microbial growth even where energy is available \cite{Higgins2024}. Our simulations assume that nutrients such as phosphorus, iron and sulfur are available in sufficient amounts for growth. Closing that carbon loop, and then propagating dissolved and particulate material through a phase-resolved ocean/plume model, would be a natural extension of the present model.

The broader takeaway of this model is inferential. Candidate communities can be treated as alternative hypotheses, propagated through the same geochemical model, and compared with the sparse observables available for ocean worlds. Ultimately, maximizing the utility of future mission data dictates that multiple ecosystem community compositions be explicitly considered during the assessment of potential biosignatures \cite{Catling2018,Cockell2024}.

\matmethods{

\subsection*{Calculating feasibility of fermentation}

As a rough maximum estimate for Enceladus's initial glycine content, we can assume cometary material as the major accretionary component, in line with its large water fraction and by analogy to Titan \cite{Miller2026,Affholder2025}. For this estimate, we assume that IOM constitutes 18\% of Enceladus's core mass.
Cometary organic matter, in particular its soluble component (SOM), is poorly constrained, however cometary organic matter likely shares a common origin with the organic matter found in carbonaceous chondrites \cite{Danger2022} and Bennu's glycine appears to originate from primordial ices akin to comets \cite{Baczynski2026}.
To get a maximum order of magnitude estimate for how IOM and glycine might scale, we take the abundance of glycine in the most glycine-rich chondrite characterized in ref. \cite{Glavin2010}, and divide by its mass fraction of insoluble carbon characterized by ref. \cite{Alexander2007}.
Doing so yields an approximate 1:100 scaling between the masses of glycine and IOM. Applying this scaling to the assumed cometary IOM inventory suggests that the mass fraction of glycine in Enceladus's core material could be as high as $0.0018$.

To compare the organic supply explored in the simulations with a possible finite source, we write the primordial glycine stock as
\begin{equation}
  m_{\rm gly}=m_{\rm core} f_{\rm IOM}({\rm GLY:IOM}),
\end{equation}
where \(m_{\rm core}=\SI{7.0e19}{\kilo\gram}\), \(f_{\rm IOM}=0.18\), and \(({\rm GLY:IOM})=0.01\) \cite{Iess2014,Park2024,Alexander2007,Alexander2012,Glavin2010}.
With these values, Enceladus may initially contain up to \SI{1.26e17}{\kilo\gram} glycine, or about \SI{1.68e18}{\mol} (using a molar mass of \SI{75}{\gram\per\mol} for glycine), suggesting that organics including glycine may be minor constituents of outer Solar system bodies rather than mere trace compounds.

The glycine-fermentation stoichiometry in Eq.~\ref{eq:glycine_fermentation} produces approximately one mole of \ch{H2} for each mole of glycine consumed. Supplying the adopted \ch{H2} escape interval, \(\Phi_{\ch{H2}}^{\rm obs}\approx(1.2\text{--}7.4)\times10^9\)\,\si{\mol\per\year}, would require \((0.9\text{--}5.5)\times10^8\)\,\si{\kilo\gram\per\year} of glycine-equivalent throughput. Dividing the stock estimate by that throughput gives a depletion time of \(\sim0.2\text{--}1.4\)\,Gyr \cite{Waite2017,Hansen2011,Hansen2020}. We compare the ecosystem model's \(\Phi_{H_2}\) and \(\Phi_{CH_4}\) outputs with Cassini separately from this stock calculation. This lifetime assumes that all of the estimated fermentable material is available to microbes. If some remains trapped in the core, the usable supply would be exhausted sooner at the same consumption rate.

Glycine is both the amino acid modeled below and a molar proxy for other fermentable substrates. We express other organics in glycine equivalents when their breakdown can be represented on a comparable \ch{H2}, short-chain-carboxylate, \ch{NH4+}, and DIC stoichiometry and energetics. Material that remains bound in the core, is poorly soluble or inaccessible, or is processed too slowly would not contribute within the vent residence time.

\subsection*{Hydrothermal plume and metabolic framework}

We use the hydrothermal mixing calculation of Affholder et al. (2021) \cite{Affholder2021}. Hydrothermal fluid enters a thin seafloor mixing layer.
The injection of hydrothermal fluid is modeled as a Gaussian function of the distance from the vent's center.
The initial upward velocity of buoyant hydrothermal fluid and seawater mix combined with the composition of the mixing layer determine the net hydrothermal export rate of volatiles.
By assumption of oceanic steady-state for \ch{H2} and \ch{CH4}, we assume that the rate of escape of these volatiles in the space plume should be comparable to the net hydrothermal release into the ocean \cite{Affholder2021}.
The equations for the source term, buoyant escape flux, steady mixing fraction, and plume average are given in SI Appendix, Methods S1.

The mixing layer may contain populations of glycine fermenters (G), hydrogenotrophic methanogens (H), and acetoclastic methanogens (A). Equation~\ref{eq:glycine_fermentation} gives the stoichiometric equation for the energy metabolism of G. The reactions for H and A, as well as thermodynamic calculations are given in SI Appendix, Methods S2. Catabolic rates depend on temperature through the Eyring equation and the temperature-dependent fraction of active enzyme \cite{Daniel2010}, and on substrate availability through a Monod term. We calculate catabolic Gibbs energies from standard thermodynamic data and local chemical concentrations \cite{Lide2005CRC}. Biomass yield is the Gibbs energy released by catabolism divided by the energy requirement for biomass synthesis \cite{Kleerebezem2010}. Biomass production depends on this yield and the catabolic rate, while maintenance \cite{Tijhuis1993} and mortality reduce net growth.

We model each radial location separately, with fluid passing through the thin mixing layer and cells remaining in place, simulating biofilm-based populations. At radial distance \(r\) (m) from the vent's center, replacement by the upward hydrothermal flow is represented by
\[
D(r)=\frac{J_c(r)}{\varepsilon\rho_o},
\]
where \(D(r)\) is the fluid-exchange rate (s\(^{-1}\)), \(J_c(r)\) is the upward convective mass-flux density (kg m\(^{-2}\) s\(^{-1}\)), \(\varepsilon\) is the mixing-layer thickness (m), and \(\rho_o\) is the water density (kg m\(^{-3}\)). The ordinary differential equations (ODEs) in Eq.~\ref{eq:main_ode_system} use the retained carbon biomass per mass of fluid,
\[
B_j(r;t)=\frac{N_j(r;t)B_j^*(r;t)}{\varepsilon\rho_o},
\]
where \(N_j\) is areal cell abundance (cells m\(^{-2}\)) and \(B_j^*\) is per-cell carbon biomass (mol C\(_x\) cell\(^{-1}\)) for metabolism \(j\). The resulting \(B_j\) has units of mol C\(_x\) kg\(^{-1}\) of fluid. We use the sign convention \(\gamma_{i,j}<0\) for substrates consumed by metabolism \(j\) and \(\gamma_{i,j}>0\) for products released to the local fluid.

Anabolic (assimilatory) uptake is restricted to the carbon source used by each guild:
\[
\alpha_{i,j}=
\begin{cases}
1/n_{C,j}, & i=s_j,\\
0, & i\ne s_j,
\end{cases}
\qquad
(s_j,n_{C,j})=
\begin{cases}
(\mathrm{DIC},1), & j=\mathrm{H},\\
(\mathrm{Gly},2), & j=\mathrm{G},\\
(\ch{CH3COO-},2), & j=\mathrm{A},
\end{cases}
\]
Here \(s_j\) is the anabolic carbon source and \(n_{C,j}\) is the number of carbon atoms in that molecule. The concentration and retained-biomass equations are
\begin{equation}
\label{eq:main_ode_system}
\begin{aligned}
\frac{\partial C_i}{\partial t}
&=D(r)\bigl(C_i^0-C_i\bigr)
+\sum_j B_j q_j^{\mathrm{cat}}\gamma_{i,j}
-\sum_j \alpha_{i,j}Y_j q_j^{\mathrm{cat}}B_j,\\
\frac{\partial B_j}{\partial t}
&=B_j\left(Y_jq_j^{\mathrm{cat}}-q_j^{\mathrm m}-d\right).
\end{aligned}
\end{equation}
In Eq.~\ref{eq:main_ode_system}, \(C_i^0\) is the abiotic inlet concentration, \(q_j^{\mathrm{cat}}\) the biomass-specific catabolic rate, \(Y_j\) the thermodynamic biomass yield, \(q_j^{\mathrm m}\) the maintenance demand, and \(d\) the shared mortality rate. The first term in the concentration equation describes fluid exchange, which drives local concentrations toward their initial abiotic steady state. The second accounts for substrates consumed and products released by catabolism, and the third for uptake of each guild's carbon source to build biomass. The biomass equation balances biomass production against maintenance and mortality losses. Because those losses are not routed into a separate dissolved reservoir, Eq.~\ref{eq:main_ode_system} is an open mixing-layer model. We use the model to compare \ch{H2} and \ch{CH4} plume fluxes among candidate communities. Closing a whole-ocean carbon budget would require an explicit return pool. Rate laws, unit conventions, and numerical details are given in SI Appendix, Methods S2--S4.

\subsection*{Community assembly and numerical simulation}

We assembled communities by integrating Eq.~\ref{eq:main_ode_system} in successive stages. Glycine fermenters were introduced first and the G-only system was integrated to equilibrium. We considered equilibrium reached when changes in dissolved concentrations and biomass satisfied the convergence criterion in SI Appendix, Methods S4. We then tested H, A, or both in the chemical environment produced by fermentation, following the GH, GA, GHA, and GAH histories. A guild was introduced only if its net growth was non-negative, with an initial biomass of \(B_j=10^{-8}\) mol C\(_x\) kg\(^{-1}\). We integrated to equilibrium again after each introduction. We identified the final community from the guilds whose local biomass exceeded \(10^{-8}\) mol C\(_x\) kg\(^{-1}\) at equilibrium (SI Appendix, Methods S4). We recorded this final composition separately from the list of guilds introduced.

We divided the mixing layer into a one-dimensional annular grid of \(n_r=100\) independent shells (no radial interaction terms in equations \ref{eq:main_ode_system}) that effectively constitute a set of ordinary differential equations (ODEs) which we integrate using an implicit backward differentiation formula (BDF) scheme\cite{Virtanen2020}. We used relative and absolute solver tolerances of \(10^{-7}\) and \(10^{-9}\), respectively. SI Appendix, Methods S4 describes how the radial positions were chosen, when each simulation stage ended, and how gas fluxes and biomass were integrated over the vent. 

\subsection*{Prior distributions}

The Standard case is our baseline for assessing fermentation-supported communities with limited abiotic \ch{H2} and \ch{CH4} inputs. The Minimal case tests how much fermentable organic material is required to reproduce the Cassini-derived gas fluxes when abiotic \ch{H2}, \ch{CH4}, and acetate concentrations are near zero. The Serpentinization-Enhanced case allows larger abiotic \ch{H2} and \ch{CH4} inputs while retaining the other Standard priors. Additional sensitivity cases examine glycine and acetate separately and together, and higher \ch{H2} inputs without abiotic methane or acetate (SI Appendix, Methods S5 and Tables S1--S2). For each principal treatment, we used 30,000 random parameter samples from the specified prior distributions. SI Appendix, Table S3 lists the biological constants and the thermodynamic, physical, flux-conversion, and numerical values.

We fixed the prior ranges before comparing simulations with Cassini. Zandanel et al. (2021) calculated global \ch{H2} production rates exceeding \(10^{12}\) mol\,yr\(^{-1}\), compared with a minimum plume-derived release rate of approximately \(10^9\) mol\,yr\(^{-1}\), suggesting that ongoing serpentinization may be nearing completion \cite{Zandanel2021}. Daval et al. (2022) estimated serpentinization lifetimes of up to approximately 500 Myr under the conditions examined and discussed limited reactions in poorly circulated regions of the core as a possible source of continued \ch{H2} production \cite{Daval2022}. Together, these studies motivate our Standard case with reduced but nonzero abiotic hydrogen inputs. We set hydrothermal upper bounds of \(10^{-3}\) mol\,kg\(^{-1}\) for \ch{H2} and \(10^{-6}\) mol\,kg\(^{-1}\) for \ch{CH4}, each 100 times lower than the corresponding bounds of Affholder et al. (2021), which we retain in the Serpentinization-Enhanced case \cite{Affholder2021}. The full prior ranges and their implications for gas supply are given in SI Appendix, Methods S5 and Table S1. The DIC and hydrothermal temperature ranges follow Affholder et al. (2021) \cite{Affholder2021}. The DIN range and the pH interval of 9--11 follow the Enceladus-like compositional scenarios of Affholder and Higgins et al. (2025) \cite{Affholder2025}; we apply the DIN interval independently to ocean water and hydrothermal fluid. Organic-substrate bounds are based on returned-sample measurements and the wider glycine-equivalent proxy described in SI Appendix, Methods S5 \cite{GlavinBennu2025,Khawaja2025,Craddock2026}. The separate oxidant-budget calculation follows Ray et al. \cite{Ray2021} (SI Appendix, Supplementary Results S2).

\subsection*{Observables and inference}

For comparison with Cassini, model concentrations were converted to annual escape fluxes of molecular hydrogen and methane. We use the INMS abundance intervals reported by Waite et al. (2017): 0.4--1.4\% for \ch{H2} and 0.1--0.3\% for \ch{CH4} by volume of the plume gas, which consists of 96--99\% water vapor \cite{Waite2017}. A later reanalysis by Peter et al. estimated a methane abundance of \(0.11\pm0.02\%\) by volume (mean \(\pm\) standard error), near the lower end of the methane interval used here \cite{Peter2024}. With a water-vapor ejection envelope of 170--300\,\si{\kilogram\per\second}, these intervals correspond to \((1.2\text{--}7.4)\times10^9\)\,\si{\mol\per\year} for \ch{H2} and \((3.0\text{--}15.8)\times10^8\)\,\si{\mol\per\year} for \ch{CH4} \cite{Waite2017,Hansen2011,Hansen2020}. For the counts and parameter summaries, we require at least one water-vapor ejection rate within the adopted range at which both modeled gas fluxes correspond to gas fractions within the observed intervals. SI Appendix, Methods S6 explains how we calculate annual plume fluxes and assess whether both gas fluxes are consistent with the Cassini measurements. We used the \ch{H2} and \ch{CH4} fluxes to compare both the amount of each gas exported and their ratio. Using \ch{CO2} or reaction quotients would require additional assumptions about ocean composition, carbonate speciation and plume transport \cite{Affholder2021}.

We also estimated ecosystem support using Approximate Bayesian Computation with Random Forests (ABC--RF) \cite{Pudlo2016}. The classifier features were \((\log_{10}\Phi_{\mathrm{H_2}}, \log_{10}\Phi_{\mathrm{CH_4}})\), and the four classes were H, GH, GA, and pooled GHA/GAH. Training used simulations in which each guild grew when introduced and the model reached equilibrium after each addition. We used a 1,500-tree classification forest to assign each \ch{H2}--\ch{CH4} flux pair to a community group \cite{Breiman2001}. A separate 1,200-tree regression forest estimates the probability of an incorrect classification, using errors from simulations excluded from classifier training. One minus this probability gives the estimated probability that the classification is correct; panel C of Figs. 3 and 4 shows its average for each classified group. Increasing the forests to 2,000 classification trees and 1,500 regression trees changed classification frequencies and mean estimated probabilities by less than one percentage point. It was run five times with fixed random seeds, equal total weight for the four classes, and simulations from the same prior sample kept together during cross-validation. Full details are given in SI Appendix, Methods S7.
}

\showmatmethods{}

\dataavail{All data, materials, and software supporting this study can be made available upon request.}

\section*{Author contributions}
\preprintcontributions

\section*{Competing interests}
\preprintdeclaration

\acknow{S.N. acknowledges support from the University of Arizona NASA Space Grant Graduate Fellowship. R.F. acknowledges support from the National Aeronautics and Space Administration through its Interdisciplinary Consortia for Astrobiology Research program (award 80NSSC21K0593), and from the U.S. National Science Foundation through its Biology Integration Institute--Implementation program (award DBI-2022070), Growing Convergence Research program (award OIA-2121155), and National Research Traineeship program (award DGE-2022055).}
\showacknow{}

\clearpage

\clearpage
\begingroup
\renewcommand*{\HyperDestNameFilter}[1]{SI-#1}
\setcounter{page}{1}
\renewcommand{\thepage}{S\arabic{page}}
\setcounter{figure}{0}
\setcounter{table}{0}
\setcounter{equation}{0}
\setcounter{footnote}{0}
\setcounter{section}{0}
\setcounter{subsection}{0}
\setcounter{subsubsection}{0}
\renewcommand{\thefigure}{S\arabic{figure}}
\renewcommand{\thetable}{S\arabic{table}}
\renewcommand{\eqref}[1]{\textup{(\ref{#1})}}
\pdfbookmark[0]{Supporting Information}{supporting-information}
\titlespacing*{\subsection}{0pt}{2.5ex plus .5ex minus .2ex}{1ex plus .2ex}

\makesititle
\section*{SI Appendix Methods}

\subsection*{Methods S1. Hydrothermal mixing and plume averaging}
\label{sec:si_methods_s1}

We follow the hydrothermal mixing calculation of Affholder et al. (2021)\ \cite{SI-Affholder2021}. In this calculation, a local hydrothermal hotspot is placed at the ocean--core interface. Hot fluid enters a shallow horizontal layer of thickness \(\varepsilon\), mixes with ocean water, and buoyantly leaves the layer. The output is a vent-scale temperature field, dilution factor, and chemical inlet for the biological model. For simplification, this model does not account for ocean circulation. 

We take the hotspot to be axisymmetric. Let \(r\) be the distance from its center. Hydrothermal fluid enters the layer at temperature \(T_f\); the surrounding ocean is at temperature \(T_o\). The total hydrothermal heat power is \(F\). The local mass-flux density of hydrothermal fluid into the layer is prescribed as a Gaussian,
\begin{equation}
J_f(r)=J_{\max}\exp\!\left[-\left(\frac{r}{c}\right)^2\right],
\quad(\mathrm{kg\,s^{-1}\,m^{-2}}),
\label{eq:si_jf}
\end{equation}
where \(J_{\max}\) (\(\mathrm{kg\,s^{-1}\,m^{-2}}\)) is the centerline hydrothermal mass-flux density and \(c\) (\(\mathrm{m}\)) is the span of the source.

The total hydrothermal mass flux entering the layer is
\begin{equation}
\dot M_{\rm HT}
=
\int_0^\infty 2\pi r J_f(r)\,\mathrm{d}r
=
\pi c^2 J_{\max}.
\label{eq:si_mdot_ht}
\end{equation}
Heat conservation gives
\begin{equation}
F=C_p(T_f-T_o)\dot M_{\rm HT},
\label{eq:si_heat_power}
\end{equation}
or
\begin{equation}
\int_0^\infty 2\pi r J_f(r)\,\mathrm{d}r
=
\frac{F}{C_p(T_f-T_o)},
\quad(\mathrm{kg\,s^{-1}}),
\label{eq:si_power_constraint}
\end{equation}
with \(C_p=4200\) J kg\(^{-1}\) K\(^{-1}\).

At a given radial position, the state of the layer is described by the hydrothermal-fluid mass fraction \(x(r,t)\) and by the mixed-fluid temperature \(T(r,t)\). The endmember \(x=1\) is undiluted hydrothermal fluid, and \(x=0\) is ocean water. Per unit area, the upward buoyant mass-flux density out of the layer is
\begin{equation}
J_c(r,t)
=
\rho_o\sqrt{2\varepsilon g\alpha\left[T(r,t)-T_o\right]},
\label{eq:si_jc_transient}
\end{equation}
where \(\rho_o=1000\) kg m\(^{-3}\), \(\varepsilon=1\) m, \(\alpha=3\times10^{-4}\) K\(^{-1}\), and \(g=0.12\) m s\(^{-2}\) \cite{SI-Turner1973,SI-Goodman2004,SI-Affholder2021}.

The mass of water per unit area in the layer is \(\varepsilon\rho_o\). Hydrothermal fluid enters at rate \(J_f\). Ocean water is entrained to supply the remainder of the upward flux, and the mixture leaves at rate \(J_c\). The hydrothermal fraction then obeys
\begin{equation}
\varepsilon\rho_o\frac{\partial x}{\partial t}
=
J_f(r)-J_c(r,t)x(r,t),
\label{eq:si_x_balance}
\end{equation}
and heat balance gives
\begin{equation}
\varepsilon\rho_o\frac{\partial T}{\partial t}
=
J_f(r)(T_f-T_o)
-
J_c(r,t)\left[T(r,t)-T_o\right].
\label{eq:si_t_balance}
\end{equation}

In the simulations, we use the steady physical mixing field. Setting \(\partial x/\partial t=\partial T/\partial t=0\) gives
\begin{equation}
J_f(r)=J_c^*(r)x^*(r),
\label{eq:si_steady_mass}
\end{equation}
and
\begin{equation}
T^*(r)-T_o=x^*(r)(T_f-T_o),
\label{eq:si_steady_temp_relation}
\end{equation}
where asterisks denote steady-state values. Substituting these relations into Eq.~\eqref{eq:si_jc_transient} gives
\begin{equation}
x^*(r)
=
\left[
\frac{J_f(r)^2}
{2\rho_o^2g\varepsilon\alpha(T_f-T_o)}
\right]^{1/3},
\label{eq:si_xstar}
\end{equation}
and
\begin{equation}
T^*(r)
=
T_o+
\left[
\frac{J_f(r)(T_f-T_o)}
{\rho_o\sqrt{2g\varepsilon\alpha}}
\right]^{2/3}.
\label{eq:si_tstar}
\end{equation}
The steady upward escape flux is then
\begin{equation}
J_c^*(r)
=
\rho_o\sqrt{2\varepsilon g\alpha\left[T^*(r)-T_o\right]}
=
\left[2\rho_o^2g\varepsilon\alpha J_f(r)(T_f-T_o)\right]^{1/3}.
\label{eq:si_jcstar}
\end{equation}

The source amplitude is fixed by requiring the hotspot center to be undiluted hydrothermal fluid:
\begin{equation}
x^*(0)=1,
\quad\mathrm{or}\quad
T^*(0)=T_f.
\label{eq:si_center_condition}
\end{equation}
This gives
\begin{equation}
J_{\max}
=
\rho_o\sqrt{2\varepsilon g\alpha(T_f-T_o)}.
\label{eq:si_jmax}
\end{equation}
The Gaussian width follows from the power constraint:
\begin{equation}
c
=
\left[
\frac{F}
{\pi C_p(T_f-T_o)J_{\max}}
\right]^{1/2}.
\label{eq:si_c_width}
\end{equation}
Equations~\eqref{eq:si_jf}--\eqref{eq:si_c_width} define the hydrothermal mixing field used for all local biological calculations.

For a chemical species \(i\), let \(C_i^f\) be its hydrothermal-fluid concentration and \(C_i^o\) its ocean concentration, both in mol kg\(^{-1}\). Before biological reactions are introduced, the mixed concentration supplied to the biological model is
\begin{equation}
C_i^0(r)
=
\frac{J_f(r)C_i^f+\left[J_c^*(r)-J_f(r)\right]C_i^o}
{J_c^*(r)}
=
x^*(r)C_i^f+\left[1-x^*(r)\right]C_i^o.
\label{eq:si_mixed_concentration}
\end{equation}
For water-like density, \(1~\mathrm{mmol\,kg^{-1}}\approx1~\mathrm{mol\,m^{-3}}\).

The biological ODEs are solved independently at each radial shell. The physical model supplies the local temperature \(T^*(r)\), the inlet concentration \(C_i^0(r)\), and the exchange rate with surrounding water,
\begin{equation}
D(r)=\frac{J_c^*(r)}{\varepsilon\rho_o},
\quad(\mathrm{s^{-1}}).
\label{eq:si_dilution_rate}
\end{equation}
Thus the physical exchange term for species \(i\) is written \(D(r)[C_i^0(r)-C_i(r)]\), with the biological source and sink terms added in Methods S3. We denote the final concentration after the local biological calculation by \(C_i(r)\). For abiotic calculations, \(C_i(r)=C_i^0(r)\).

We average model outputs over the water leaving the mixing layer. With \(2\pi r\,\mathrm{d}r\) as the axisymmetric area element, the flux-weighted concentration of species \(i\) in the exported oceanic plume water is
\begin{equation}
C_i^{\rm OP}
=
\frac{
\displaystyle\int_0^\infty 2\pi r J_c^*(r)C_i(r)\,\mathrm{d}r
}{
\displaystyle\int_0^\infty 2\pi r J_c^*(r)\,\mathrm{d}r
},
\quad(\mathrm{mol\,kg^{-1}}).
\label{eq:CiOP}
\end{equation}
The corresponding instantaneous escape flux from the modeled vent layer is
\begin{equation}
\Phi_i
=
\int_0^\infty 2\pi r J_c^*(r)C_i(r)\,\mathrm{d}r,
\quad(\mathrm{mol\,s^{-1}}).
\label{eq:si_escape_flux}
\end{equation}
Since \(J_c^*\) is a mass-flux density and \(C_i\) is in mol kg\(^{-1}\), the product \(J_c^*C_i\) has units mol s\(^{-1}\) m\(^{-2}\) . We therefore do not apply an additional factor of \(1/\rho_o\). Annual fluxes are obtained by multiplying by \(t_{\rm yr}=365\times24\times3600\) s yr\(^{-1}\), as described in Methods S6.

Molecular diffusion is not included in this physical model; transport is assumed to be dominated by mixing-layer entrainment and buoyant export \cite{SI-Turner1973,SI-Goodman2004}. We compare the fluxes and flux-weighted compositions leaving the seafloor mixing layer with the plume-derived constraints in Methods S6. As in Affholder et al.\ \cite{SI-Affholder2021}, dilution during ascent through the ocean is not resolved.

\subsection*{Methods S2. Catabolic energetics and temperature-dependent rates}
\label{sec:si_methods_s2}

We use the framework of Affholder et al.\ \cite{SI-Affholder2021} to link microbial catabolism, enzyme kinetics, maintenance, and anabolism under the local temperature and substrate conditions. Each catabolic reaction \(j\) is written over chemical species \(\{A_k\}\) as
\begin{equation}
\sum_k \nu_{k,j} A_k = 0,
\end{equation}
where \(\nu_{k,j}<0\) for reactants and \(\nu_{k,j}>0\) for products. Reactions are normalized per mole of electron donor (eD). The reactions below are displayed as ordinary chemical equations, with positive stoichiometric magnitudes on both sides; in the numerical reaction vector they follow the signed convention above. After glycine fermenters, whose stoichiometry is given in main-text Eq. 1, have been tested for viability, the following methanogenic metabolisms can be introduced:
\[
\text{Hydrogenotrophic methanogenesis:}\quad
\ch{H2 + 1/4 HCO3- + 1/4 H+}
\longrightarrow
\ch{1/4 CH4 + 3/4 H2O}.
\]
\[
\text{Acetoclastic methanogenesis:}\quad
\ch{CH3COO- + H2O}
\longrightarrow
\ch{CH4 + HCO3-}.
\]

Because stoichiometries are normalized per mole of electron donor, reaction energies and rates are expressed per mol$_{\mathrm{eD}}\). The Gibbs free-energy change for each catabolic reaction (J mol\(_{\mathrm{eD}}^{-1}\)) is evaluated at the local temperature \(T\) (K):
\begin{equation}
\Delta G_{\rm cat}(T,Q_{\rm cat})=\Delta G^0_{\rm cat}(T)+RT\ln Q_{\rm cat},
\end{equation}
where \(Q_{\rm cat}=\prod_k a_k^{\nu_{k,j}}\), \(a_k\) is the activity of species \(k\), and \(\nu_{k,j}\) follows the signed reaction-vector convention. At unit activity, the standard free energy is written
\begin{equation}
\Delta G^0_{\rm cat}(T)=\Delta G^0_{S}\,\frac{T}{T_S}+\Delta H^0_{S}\,\frac{T_S - T}{T_S},
\end{equation}
where \(T_S = 298\) K. The quantities \(\Delta G^0_S\) and \(\Delta H^0_S\) are the reaction-specific standard Gibbs free energies and standard enthalpies at \(T_S\), normalized per mole of electron donor from CRC thermodynamic data and listed in SI Appendix, Table S3 \cite{SI-Lide2005CRC}.

We apply acid--base corrections in reaction-specific \(\Delta G^0_{\rm cat}(T,\mathrm{pH})\) functions, following thermodynamic state-analysis treatments of pH-dependent reaction energetics \cite{SI-Kleerebezem2010}. Gas--solution corrections are also applied reaction by reaction. We define
\begin{equation}
\begin{aligned}
\chi_{\rm DIC}(T,\mathrm{pH})&=RT\ln\!\frac{K_1(T)\,[\mathrm{H}^+]}{K_1(T)\,[\mathrm{H}^+] + [\mathrm{H}^+]^2 + K_1(T)\,K_2(T)},\\
\chi_{\rm DIN}(T,\mathrm{pH})&=-RT\ln\!\Bigl(1 + \frac{K_a(T)}{[\mathrm{H}^+]}\Bigr),\\
\chi_{\rm H}(T,\mathrm{pH};\nu_H)&=\nu_HRT\ln[\mathrm{H}^+],\\
\chi_{\rm gas,H_2/CH_4}(T)&=-RT\left[-\ln K_H^{H_2}+\frac{1}{4}\ln K_H^{CH_4}\right],
\end{aligned}
\end{equation}
where \([\mathrm{H}^+]=10^{-\mathrm{pH}}\), \(K_a\) is the NH\(_4^+\)/NH\(_3\) dissociation constant \cite{SI-Erickson1985,SI-Emerson1975}, and \(K_1,K_2\) are the first and second dissociation constants of carbonic acid \cite{SI-ClarkFritz1997,SI-HeMorse1993}. The gas term uses the Henry-law concentration/partial-pressure convention \(K_H=C_i/P_i\). In the present implementation \(K_H^{H_2}=7.8\times10^{-4}\) and \(K_H^{CH_4}=1.4\times10^{-3}\) are fixed constants. DIC speciation enters hydrogenotrophic methanogenesis through the \(-0.25\,\chi_{\rm DIC}\) term. The correction terms are therefore \(\Delta G^0_{\rm gly}=\Delta G^0_{\rm gly,base}+\chi_{\rm DIN}+\chi_{\rm DIC}+\chi_{\rm H}(\nu_H=0.5)\), \(\Delta G^0_{\rm H}=\Delta G^0_{\rm H,base}+\chi_{\rm gas,H_2/CH_4}-0.25\,\chi_{\rm DIC}+\chi_{\rm H}(\nu_H=-0.25)\), and \(\Delta G^0_{\rm A}=\Delta G^0_{\rm A,base}+\chi_{\rm DIC}\) for acetoclastic methanogenesis written in bicarbonate form. The acetate pool is represented as \ch{CH3COO-}. We do not apply a separate acetate/acetic-acid correction (because acetate dominates at pH 9--11) and is treated as one alkaline acetate pool. Likewise, glycine is represented as a single aqueous glycine-equivalent substrate, so no separate glycine/glycinate correction is added beyond the DIC, DIN, and proton terms above. Concentrations are bounded below by zero during integration. Reaction quotients are then evaluated from the resulting concentrations, excluding species whose concentration is zero.

Following Daniel et al.\ \cite{SI-Daniel2010}, we model the per-enzyme catalytic rate with an Eyring / transition-state-theory expression for catalysis by the activated enzyme (\(k_{\mathrm{cat}},\ \mathrm{mol}_{\mathrm{eD}}\ \mathrm{s}^{-1}\,\mathrm{mol}_{E_{\mathrm{act}}}^{-1}\)). We also use an equilibrium law for the ratio of inactivated to activated enzyme \(K_{\mathrm{eq}}=E_{\mathrm{inact}}/E_{\mathrm{act}}\):
\begin{equation}
\begin{aligned}
v_{\mathrm{max}}&= \frac{k_{\mathrm{cat}}}{1 + K_{\mathrm{eq}}},\\
k_{\mathrm{cat}}(T)&= \frac{k_B\,T}{h}\,\exp\!\Bigl(-\frac{\Delta G^{\ddagger}_{\mathrm{cat}}}{R\,T}\Bigr),\\
K_{\mathrm{eq}}&= \exp\!\Bigl[\frac{\Delta H_{\mathrm{eq}}}{R}\Bigl(\tfrac{1}{T_{\mathrm{eq}}}-\tfrac{1}{T}\Bigr)\Bigr].
\end{aligned}
\end{equation}
Here \(T\) and \(T_{\mathrm{eq}}\) are absolute temperatures in Kelvin. In the reported model \(T_{\mathrm{eq}}=363.15\) K (90~\si{\celsius}), and \(\Delta G^{\ddagger}_{\mathrm{cat}}=\)\SI{72}{\kilo\joule\per\mol} is a fixed activation barrier shared by the instantiated metabolisms. This activation barrier is distinct from the reaction free energy \(\Delta G_{\rm cat}(T,Q_{\rm cat})\).

Scaling this rate by the enzyme-to-biomass ratio \(\tau\) and by a Monod term for the limiting-substrate concentration \(C\) gives
\begin{equation}
q_{\rm cat}=\tau\,v_{\rm max}\,\frac{C}{C + K_m},
\end{equation}
where \(K_m\) is the half-saturation constant, expressed on the same mass basis as \(C\) (mol\,kg\(^{-1}\)). Under fixed water-like density, this is equivalent to the density-normalized aqueous update used in the numerical formulation.

Part of the catabolic energy flux is assigned to maintenance. We use the empirical maintenance-energy function of Tijhuis et al., in the form used by Affholder et al.\ \cite{SI-Tijhuis1993,SI-Affholder2021}, which has the form of an Arrhenius expression for biomass destabilization. In the units used here, the reference value \(84~\mathrm{kJ\,C\text{-}mol^{-1}\,day^{-1}}\) at \(T_S=298\) K is rewritten in \(\mathrm{J\,s^{-1}\,mol^{-1}_{C_x}}\). This introduces the conversion factor \(\tfrac{1000}{86400}=\tfrac{5}{432}\) (kJ\(\to\)J and day\(\to\)s).
\begin{equation}
e_m(T)=84\exp\!\Bigl[\frac{69400}{R}\bigl(\tfrac{1}{T_S}-\tfrac{1}{T}\bigr)\Bigr]\cdot\left(\frac{5}{432}\right)
\quad(\mathrm{J\,s^{-1}\,mol^{-1}_{C_x}})
\end{equation}
Each metabolism is assigned an effective anabolic energy cost \(E_{\rm ana,j}\) (J\,mol\(^{-1}_{C_x}\)). Hydrogenotrophic methanogenesis is treated as autotrophic growth on DIC with \ch{H2} as a strong electron donor. For this case we use \(E_{\rm ana,H}=9.86\times10^5~\mathrm{J\,mol^{-1}_{C_x}}\), the no-reversed-electron-transfer \(\Delta G_{\rm diss}(\mathrm{CO_2})\) value of Kleerebezem and Van Loosdrecht \cite{SI-Kleerebezem2010}. For heterotrophic glycine fermenters and acetoclastic methanogens, we use the empirical organic-carbon-source approximation \(E_{\rm ana,j} = \Delta G_{\rm diss}(N_{C,j},\gamma_j)\), with
\begin{equation}
\Delta G_{\rm diss}(N_{C,j},\gamma_j)=\left[200+18(6-N_{C,j})^{1.8}+\exp\!\left(((-0.2-\gamma_j)^2)^{0.16}(3.6+0.4N_{C,j})\right)\right]\times10^3,
\end{equation}
\(N_{C,j}\) is the number of carbon atoms in the organic carbon source, and \(\gamma_j\) is its average carbon oxidation state. Glycine fermentation uses \(N_C=2,\gamma=1\), giving \(E_{\rm ana,G}\approx5.24\times10^5~\mathrm{J\,mol^{-1}_{C_x}}\). Acetoclastic methanogenesis uses acetate with \(N_C=2,\gamma=0\), giving \(E_{\rm ana,A}\approx4.32\times10^5~\mathrm{J\,mol^{-1}_{C_x}}\). The corresponding gross biomass yield and maintenance-loss term are
\begin{equation}
Y_j=-\frac{\Delta G_{\rm cat}}{E_{\rm ana,j}}
\quad(\mathrm{mol\;C_x\;mol\;eD}^{-1}),
\qquad
q_j^{\mathrm{m}}=\frac{e_m(T)}{E_{\rm ana,j}}
\quad(\mathrm{s}^{-1}).
\end{equation}
Thus \(Y_j\) varies with the local thermodynamic driving force. The term \(q_j^{\mathrm{m}}\) varies with temperature through \(e_m(T)\) and enters directly as a biomass-specific loss term in the growth equation below.

\noindent The baseline mortality used throughout is \(d=1.0\times10^{-6}\ {\rm s}^{-1}\).

\subsection*{Methods S3. Multispecies substrate and biomass dynamics}
\label{sec:si_methods_s3}

We assume that the hypothetical ecosystem occupies a local hydrothermal environment. The local chemical conditions \(C_i\) are set by ocean and hydrothermal-fluid compositions, temperature, and flow-through substrate turnover in the idealized mixing model described above. If a metabolism is viable, it changes those local conditions in a metabolism-specific way \cite{SI-Cockell2024}. The equilibrium reached after this biological modification is then passed to the plume-flux calculation.

In each radial shell around the vent, we represent dissolved substrates as a continuously mixed, flow-through aqueous reservoir. Biomass, in contrast, is assumed to remain locally retained within the mixing layer over the timescale of the shell calculation. Dissolved substrates \(C_{i}(r;t)\) are reported on a mass basis (mol kg\(^{-1}\)) and are supplied and removed by physical flow at rate
\begin{equation}
D(r)\equiv \frac{J_c(r)}{\varepsilon\,\rho_o}\quad(\mathrm{s}^{-1}),
\end{equation}
which converts the buoyant flux density \(J_c\) (kg m\(^{-2}\) s\(^{-1}\)) into a per-second turnover rate by dividing by the mass of water \(\varepsilon\,\rho_o\) (kg m\(^{-2}\)) in the layer. In the numerical formulation, reactive source and sink terms are evaluated in a density-normalized aqueous concentration update with fixed \(\rho_o\). Concentrations are reported externally on a mol\,kg\(^{-1}\) basis for consistency with priors, tables, and plume comparisons. The biomass state is defined as
\[
B_j(r;t)\equiv \frac{N_j(r;t)\,B_j^*(r;t)}{\varepsilon\,\rho_o},
\]
where \(N_j\) is the local areal abundance of metabolism \(j\) (number m\(^{-2}\)), and \(B_j^*\) is its per-cell carbon biomass (mol C\(_x\) cell\(^{-1}\)). The resulting \(B_j\) has units mol C\(_x\) kg\(^{-1}\), and is the collapsed biomass variable used in the numerical update. Each metabolism \(j\) consumes or produces substrate \(i\) according to its signed stoichiometric coefficient \(\gamma_{i,j}\equiv\nu_{i,j}\) at biomass-specific catabolic rate \(q_{j}^{\mathrm{cat}}\) (mol eD mol C\(_x^{-1}\) s\(^{-1}\)). Values \(\gamma_{i,j}<0\) denote consumption, and values \(\gamma_{i,j}>0\) denote production. A fraction of the catabolic energy is converted into biomass at yield
\begin{equation}
Y_{j}=-\frac{\Delta G_{\rm cat}}{E_{\rm ana,j}}\quad(\mathrm{mol\;C_x\;mol\;eD}^{-1}),
\end{equation}
while maintenance enters as the biomass-specific loss term \(q_{j}^{\mathrm{m}}=e_m(T)/E_{\rm ana,j}\) (s\(^{-1}\)). For all radii \(0<r\le R_{\max}\), the coupled system is then
\begin{equation}
\alpha_{i,j}=\begin{cases}
1/n_{C,j}, & i=s_j,\\
0, & i\ne s_j,
\end{cases}
\qquad
(s_j,n_{C,j})=\begin{cases}
(\mathrm{DIC},1), & j=\mathrm{H},\\
(\mathrm{Gly},2), & j=\mathrm{G},\\
(\mathrm{CH_3COO^-},2), & j=\mathrm{A},
\end{cases}
\end{equation}
where \(s_j\) is the anabolic carbon source and \(n_{C,j}\) is the number of carbon atoms per molecule of that source:
\begin{equation}
\begin{aligned}
\frac{\partial C_{i}}{\partial t}
&= D(r)\bigl(C^0_{i}-C_{i}\bigr)+\sum_{j}B_{j}\,q_{j}^{\mathrm{cat}}\,\gamma_{i,j}-\sum_j \alpha_{i,j}Y_{j}\,q_{j}^{\mathrm{cat}}\,B_{j},\\
\frac{\partial B_{j}}{\partial t}
&= B_{j}\Bigl(Y_{j}\,q_{j}^{\mathrm{cat}}- q_{j}^{\mathrm{m}}- d\Bigr).
\end{aligned}
\end{equation}
The first term in the substrate equation gives wash-in of fresh fluid at concentration \(C^0_i\) and wash-out of resident \(C_i\). The second term gives the net microbial transformation of each substrate by catabolism. The third term removes the substrate used to build biomass: DIC for hydrogenotrophic methanogens, glycine for fermenters, and acetate for acetoclastic methanogens. Dividing by \(n_{C,j}\), the number of carbon atoms per substrate molecule, converts the yield from moles of biomass carbon to moles of substrate.

This is an open-system approximation. The biomass variable \(B_j=N_jB_j^*/(\varepsilon\rho_o)\) is the retained, density-normalized numerical biomass state. Maintenance and mortality reduce modeled biomass, but we do not route that carbon into an explicit dissolved receiving pool. We assume that cells are attached or otherwise retained in the mixing layer while dissolved substrates flow through it. The biomass equation therefore includes no loss through fluid flow. The model is therefore used for comparative plume-flux inference, not for a closed Enceladus carbon budget or an absolute biomass inventory.

For ecological assembly, we introduce a metabolism only when it is viable under the local conditions present at that stage. We evaluate the net specific growth rate
\begin{equation}
\mu_j \;=\; Y_j\,q_j^{\mathrm{cat}} \;-\; q_j^{\mathrm{m}} \;-\; d.
\end{equation}
If \(\mu_j\ge 0\), metabolism \(j\) is seeded at a small numerical inoculum (\(B_j=10^{-8}\ \mathrm{mol\,C_x\,kg^{-1}}\)). If \(\mu_j<0\), it is not inoculated at that stage. The same scalar mortality \(d=1.0\times10^{-6}\ \mathrm{s^{-1}}\) is applied to all metabolisms. After each introduction, we integrate the ODE system until it meets the equilibration criterion (Methods S4). We then test whether the next guild in the prescribed order has non-negative net growth under the resulting chemical conditions \cite{SI-DieckmannLaw1996}.

\subsection*{Methods S4. Numerical simulation and implementation}
\label{sec:si_methods_s4}

We discretize the vent as a one-dimensional radial grid of concentric annular cells. For each parameter sample, we set the grid's extent from the hydrothermal mixing profile so that it covers both the region influenced by the vent and surrounding fluid approaching ocean conditions. The hydrothermal source is the Gaussian \(J_f(r)=J_{\max}\exp[-(r/c)^2]\), with \(J_{\max}\) and \(c\) set by the centerline-temperature and total-power closure above. From the steady-state expression for \(x^*(r)\), the centerline hydrothermal mixing ratio is
\[
x_0\equiv x^*(0)=\left[\frac{J_{\max}^2}{2\rho_o^2 g\,\varepsilon\,\alpha\,(T_f-T_o)}\right]^{1/3},
\]
and therefore
\[
x^*(r)=x_0\exp\!\left[-\frac{2}{3}\left(\frac{r}{c}\right)^2\right].
\]
For a chosen target hydrothermal fraction \(x_{\mathrm{target}}\), the transition radius is obtained by solving \(x^*(r)=x_{\mathrm{target}}\):
\[
r_{\mathrm{transition}}=c\sqrt{\frac{3}{2}\ln\!\left(\frac{x_0}{x_{\mathrm{target}}}\right)}.
\]
We then set \(R_{\max}=\sigma\,r_{\mathrm{transition}}\), so that \(\sigma\) controls how far into the far field the integration extends.

In the reported simulations, \(n_r=100\), \(x_{\mathrm{target}}=10^{-5}\), and \(\sigma=3\). We use scaled spacing, with approximately 85\% of shells allocated to \([0,r_{\mathrm{transition}}]\) and the remaining 15\% to \([r_{\mathrm{transition}},R_{\max}]\). This places more resolution near the vent, where gradients are steepest. Shell centers are edge midpoints, and shell widths \(\Delta r_k\) are differences between adjacent edges.

At each shell center \(r_k\), we compute \(J_f(r_k)\), \(T^*(r_k)\), \(J_c(r_k)\), and the abiotic inlet concentrations \(C_i^0(r_k)\). An independent local ecosystem simulation is then run for that shell, with dilution rate \(D(r_k)=J_c(r_k)/(\varepsilon\rho_o)\). The stiff ODEs are integrated using a backward differentiation formula (BDF) method with adaptive internal timesteps, a relative tolerance of \(10^{-7}\), and an absolute tolerance of \(10^{-9}\) \cite{SI-Virtanen2020}. During integration, non-physical negative concentrations or biomasses are bounded below by zero. Newly introduced biomasses are initialized at \(B_j=10^{-8}\), and thermodynamic quantities are evaluated from the resulting non-negative state using the numerical convention described above.

For the ABC--RF and sensitivity analyses, all required stages had to reach equilibrium at at least one common radial location. At that location, each subsequent guild was introduced into the equilibrated chemical environment produced by the preceding community. After each successful inoculation, integration proceeds in successive windows of duration \(\Delta t_{\mathrm{check}}=2\times10^7\) s. Each stage ends when the convergence criterion is met or its maximum duration is reached. Equilibrium is assessed by comparing the state vector \(y=(C_i,B_j)\) between successive checks using the component-wise tolerance
\[
|y^{(n)}-y^{(n-1)}| \le a_{\mathrm{eq}} + r_{\mathrm{eq}}\max\!\bigl(|y^{(n)}|,|y^{(n-1)}|\bigr)
\]
with \(a_{\mathrm{eq}}=10^{-9}\) and \(r_{\mathrm{eq}}=10^{-7}\). Each stage was subject to a minimum duration of \(10^6\) s and an initial maximum of \(7\times10^8\) s. Targeted reruns used the original parameter values and increased the maximum stage duration to \(1.4\times10^9\), \(2.8\times10^9\), and \(5.6\times10^9\) s. At the largest limit, locations that had not reached equilibrium could contribute no more than 1\% of either gas flux, and each gas flux had to change by no more than 0.01 dex between the last two durations. After numerical checks and exclusions, the analyses contained 29,970 Minimal, 29,994 Standard, and 29,942 Serpentinization-Enhanced parameter samples.

Plume fluxes and integrated biomass are evaluated with annular shell weights, \(\Delta A_k = 2\pi r_k\,\Delta r_k\), on the (possibly nonuniform) radial grid:
\begin{equation}
\Phi_i \approx \sum_k J_c(r_k)\,C_i(r_k)\,\Delta A_k,\qquad
\Phi_{\mathrm{conv}} \approx \sum_k J_c(r_k)\,\Delta A_k.
\end{equation}
The same raw shell weights are used for the biomass summaries reported in the analysis:
\[
\mathcal{B}_j\equiv\sum_k B_j(r_k)\Delta A_k.
\]
Here \(B_j=N_jB_j^*/(\varepsilon\rho_o)\) is the collapsed numerical biomass variable with units mol C\(_x\) kg\(^{-1}\). Thus \(\mathcal{B}_j\) is an area-weighted summary variable with units mol C\(_x\) m\(^2\) kg\(^{-1}\). With \(J_c\) in kg m\(^{-2}\) s\(^{-1}\) and \(C_i\) in mol\,kg\(^{-1}\), each flux term \(J_c(r_k)C_i(r_k)\Delta A_k\) has units mol\,s\(^{-1}\). For integrated time series, local trajectories are interpolated onto a common time vector and then area-weighted by \(\Delta A_k\). Before a local trajectory starts, its initial value is used; after it ends, its final value is used.

For the glycine- and acetate-supply comparisons in main-text Fig. 2 and Fig. S4, we define the annual hydrothermal mass throughput as
\[
M_{\mathrm{HT},yr}\equiv \dot M\,t_{\rm yr}=\frac{F\,t_{\rm yr}}{C_p(T_f-T_o)}
\]
and use the corresponding supply metric \(M_{\mathrm{HT},yr}[X]_f\), where \([X]_f\) is the hydrothermal-fluid concentration of the plotted substrate. With \(M_{\mathrm{HT},yr}\) in kg\,yr\(^{-1}\) and \([X]_f\) in mol\,kg\(^{-1}\), this metric has units mol\,yr\(^{-1}\).

To explore prior uncertainty, we generate 30{,}000 random parameter samples from the specified prior distributions for each case assessed in the Results and ABC comparisons. For the Minimal benchmark, the non-negligible inputs are \(([\mathrm{Gly}]_f,\allowbreak[\mathrm{DIC}]_f,\allowbreak[\mathrm{DIC}]_o,\allowbreak[\mathrm{DIN}]_{o,f},\allowbreak T_f,\allowbreak\mathrm{pH})\); all other donor and acceptor pools are fixed at their Minimal values (SI Appendix, Table S1). For the Standard and Serpentinization-Enhanced analyses, the varied prior vector is the full set reported in SI Appendix, Table S1. Thermodynamic and kinetic evaluations, including Gibbs energies, enzymatic rates, maintenance costs, and anabolic costs, are computationally accelerated to allow repeated evaluations at this scale \cite{SI-Lam2015}. The resulting simulations were used for the interval counts and ABC--RF analysis. A guild was included in the final community when its local biomass exceeded \(10^{-8}(1+10^{-6})\) mol C\(_x\) kg\(^{-1}\) at at least one radial location whose final integration stage reached equilibrium. A simulation was classified as inhabited when at least one guild met this condition.

\subsection*{Methods S5. Cases assessed and their priors}
\label{sec:si_methods_s5}

We evaluate three cases in the same hydrothermal--ocean mixing framework, with concentrations expressed in mol\,kg$^{-1}$: Minimal, Standard, and Serpentinization-Enhanced. Additional targeted sensitivity treatments are given in SI Appendix, Table S2. These include the high-\ch{H2} with glycine-only case, the glycine-only case, the acetate-only case, and organics-only glycine+acetate case.

\emph{Minimal case.} In this case, H$_2$, CH$_4$, and acetate are fixed at negligible background values in both ocean and vent fluids. DIC, DIN, \(T_f\), and pH retain the ranges used in the other treatments. Hydrothermal glycine is then the only introduced source that can generate \ch{H2} and acetate through main-text Eq.~(1). This gives the range of $[\mathrm{Gly}]_f$ required to reach the Cassini-derived flux intervals and the biological \ch{H2} output obtainable with negligible abiotic \ch{H2}, \ch{CH4}, and acetate inputs.

\emph{Standard (baseline, without assuming sustained global serpentinization).} The Standard case tests whether the modeled communities can reproduce both Cassini-derived gas fluxes when abiotic \ch{H2} and \ch{CH4} inputs are limited. These inputs are specified as concentrations in hydrothermal fluid and ocean water, held constant during each simulation. The gas concentrations, glycine, acetate, DIC, DIN, temperature, and pH are sampled from the priors in Table S1, whose sources and assumptions are described below.

\emph{Serpentinization-Enhanced case.} We broaden the hydrothermal and oceanic \ch{H2} priors and allow a higher abiotic \ch{CH4} background than in the Standard case. All other priors follow the Standard case, so the comparison tests the effect of these higher abiotic inputs on ecosystem composition.

The prior families in Tables S1 and S2 were set before applying the Cassini \ch{H2}/\ch{CH4} flux window. The Standard case spans hydrothermal concentrations of \(10^{-9}\)--\(10^{-3}\) mol\,kg\(^{-1}\) for \ch{H2} and \(10^{-12}\)--\(10^{-6}\) mol\,kg\(^{-1}\) for \ch{CH4}. Oceanic concentrations span \(10^{-10}\)--\(3\times10^{-7}\) mol\,kg\(^{-1}\) for \ch{H2} and \(10^{-12}\)--\(10^{-7}\) mol\,kg\(^{-1}\) for \ch{CH4}. These ranges test reduced abiotic supply.\par Their physical scale follows from the hydrothermal heat balance in Eq.~\eqref{eq:si_heat_power}. With \(F=5\) GW, \(T_o=275\) K, and \(T_f=300\)--620 K, the annual hydrothermal throughput is approximately \(1.1\times10^{11}\)--\(1.5\times10^{12}\) kg. At the Standard hydrothermal concentration upper limits, this supplies \(1.1\times10^8\)--\(1.5\times10^9\) mol of \ch{H2} and \(1.1\times10^5\)--\(1.5\times10^6\) mol of \ch{CH4} per year. Including entrained ocean water at its upper concentration limits gives maximum modeled abiotic exports from the mixing layer of approximately \(1.5\times10^9\) mol\,yr\(^{-1}\) for \ch{H2} and \(1.8\times10^6\) mol\,yr\(^{-1}\) for \ch{CH4}, before biological processing.\par The maximum abiotic hydrogen flux is of order \(10^9\) mol\,yr\(^{-1}\), compared with production rates exceeding \(10^{12}\) mol\,yr\(^{-1}\) in Zandanel et al.'s simulations \cite{SI-Zandanel2021}. It reaches the lower end of the adopted Cassini hydrogen range. The maximum abiotic methane export is only about 0.6\% of the minimum Cassini-derived methane flux. Thus, the Standard case permits substantial abiotic hydrogen but little abiotic methane.\par The oceanic and hydrothermal DIC ranges and the hydrothermal temperature prior follow Affholder et al. \cite{SI-Affholder2021}. We use a log-uniform DIN prior spanning \(7.05\times10^{-9}\)--\(2.87\times10^{-3}\) mol\,kg\(^{-1}\), the range of oceanic total ammonia in the Enceladus-like compositional scenarios of Affholder and Higgins et al. (2025) \cite{SI-Affholder2025}. Applying this range independently to ocean water and hydrothermal fluid is an assumption of this study. We also sample pH uniformly between 9 and 11, spanning the pH values of those scenarios. The central staged-community runs keep oxidant- and reductant-equivalent pools near zero; Supplementary Results S2 gives the separate Ray et al.\ oxidant-budget estimate \cite{SI-Ray2021}.

For organics, the starting point is returned carbonaceous material. Glycine uses Bennu and Ryugu amino-acid measurements \cite{SI-GlavinBennu2025,SI-Naraoka2023Sci,SI-Parker2023,SI-Potiszil2023}, together with free and bound/precursor extraction pools \cite{SI-Glavin2010}. We extend the glycine-equivalent prior above the estimated concentration of glycine leached from rock to allow for other sources of fermentable organic material in the vent fluid. It represents a pool that could include additional leached soluble organics, amino acid products formed during hydrothermal chemistry, and organic matter generated during an earlier autotrophic or methanogenic phase when water--rock \ch{H2} production was stronger. Steel et al.\ provide the Enceladus link between hydrothermal \ch{H2}, amino-acid synthesis, and a waning serpentinization front \cite{SI-Steel2017}. Affholder et al. (2022)\ provide the corresponding methanogenic biomass and productivity calculation, including glycine-bearing organic output \cite{SI-Affholder2022}. The prior also refers to organic compounds in freshly ejected Enceladus plume grains \cite{SI-Khawaja2025} and laboratory simulations of organic synthesis in Enceladus \cite{SI-Craddock2026}. Affholder and Higgins et al. (2025)\ describe an earlier glycine-fermentation model for Titan \cite{SI-Affholder2025}. The acetate prior uses acetic acid measured in Bennu samples. Other small carboxylates are excluded from the estimate \cite{SI-GlavinBennu2025}. Tables S1 and S2 give the brackets used in the simulations. The glycine and acetate conversions follow.

\textbf{Vent fluid concentration of glycine and acetate from analog rock abundance.} 
\\Bennu and Ryugu abundances are rock-normalized, usually nmol\,g\(^{-1}\). The vent ODEs use aqueous mol\,kg\(^{-1}\). We convert with a water/rock ratio \(W/R\), an assumed availability fraction \(f_{\mathrm{avail}}\), and a release factor \(f_{\mathrm{leach}}\).

\emph{Rock inventories.} Bennu hot-water extracts contain approximately \(13{,}600\) nmol\,g\(^{-1}\) ammonia, \(33\) nmol\,g\(^{-1}\) free glycine, \(1{,}436\) nmol\,g\(^{-1}\) acetic acid, and \(4{,}106\) nmol\,g\(^{-1}\) formic acid. Acid hydrolysis of the hot-water extract gives \(44\) nmol\,g\(^{-1}\) total glycine, including the free pool \cite{SI-GlavinBennu2025}. They also contain nine C1--C7 monocarboxylic acids and two dicarboxylic acids. Ribose, glucose, and other Bennu sugars are used here as context for preservation of labile organics in the parent material \cite{SI-Furukawa2025}. The glycine prior itself uses the amino-acid measurements above. For CI/CM material, hot-water extraction and acid hydrolysis give different free and bound/precursor pools \cite{SI-Glavin2010}.

\emph{Estimating dissolved concentrations.} For species \(X\), \(C_{\mathrm{rock}}^{X}\) denotes the abundance in rock, in mol\,kg\(^{-1}\) rock. Contact 1 kg of rock with \(W/R\) kg of water, and release the fraction \(f_{\mathrm{avail}}^{X}f_{\mathrm{leach}}^{X}\). The resulting fluid concentration is
\begin{equation}
\label{eq:wr_map_prior}
[X]_f \;\approx\; \frac{C_{\mathrm{rock}}^{X}\, f_{\mathrm{avail}}^{X}\, f_{\mathrm{leach}}^{X}}{W/R}\quad(\mathrm{mol\,kg^{-1}\ water}),
\end{equation}
Here \(f_{\mathrm{avail}}^{X}\) is the fraction of the measured inventory assumed to become available for release. For glycine, a value of 1 includes the total pool measured after hydrolysis of the hot-water extract. The factor \(f_{\mathrm{leach}}^{X}\) accounts for incomplete contact, finite residence time, and resorption during mobilization into fluid. The water/rock notation follows Enceladus-like leaching calculations \cite{SI-Zolotov2007,SI-Glein2015}. Free amino acids are those measured in the hot-water extract without acid hydrolysis. The bound/precursor pool refers to the additional amino acids recovered after acid hydrolysis of that extract \cite{SI-Glavin2010,SI-GlavinBennu2025,SI-Potiszil2023}. Concentrations are reported in mol\,kg$^{-1}$; for dilute aqueous solutions with \(\rho \approx 1\)~kg\,L\(^{-1}\), \(1~\mathrm{mmol\,kg^{-1}}\) equals \(1~\mathrm{mol\,m^{-3}}\) numerically.

\emph{Glycine calculations.} We use \(C_{\mathrm{rock}}^{\mathrm{Gly}}\in[10,44]\)~nmol\,g\(^{-1}\) as a glycine-rich returned-sample bracket. The upper value is the total glycine measured after hydrolysis of the Bennu hot-water extract, including the free pool. The lower value is a practical glycine-rich starting point for the calculation. The lower hydrothermal glycine concentration estimate is then set by dilution and leachability in Eq.~\eqref{eq:wr_map_prior}. In molar units, the bracket is \((1.0\text{--}4.4)\times10^{-5}\)~mol\,kg\(^{-1}\) rock \cite{SI-GlavinBennu2025,SI-Parker2023,SI-Naraoka2023Sci}. We assume \(f_{\mathrm{avail}}^{\mathrm{Gly}}\in[0.3,1]\) and \(f_{\mathrm{leach}}^{\mathrm{Gly}}\in[0.8,1]\), with \(W/R=10\) for the lower concentration estimate and \(W/R=0.3\) for the upper estimate. These factors and dilution ratios are choices for this calculation, rather than measured intervals.

\emph{Lower concentration estimate.} With \(C_{\mathrm{rock}}^{\mathrm{Gly}}=1.0\times10^{-5}\), \(f_{\mathrm{avail}}^{\mathrm{Gly}}=0.3\), \(f_{\mathrm{leach}}^{\mathrm{Gly}}=0.8\), and \(W/R=10\):
\[
[\mathrm{Gly}]_f
\approx
\frac{1.0\times10^{-5}\times0.24}{10}
=
2.4\times10^{-7}\ \mathrm{mol\,kg^{-1}}
\quad
(=0.24\ \mu\mathrm{mol\,kg^{-1}}).
\]

\emph{Upper concentration estimate.} We assume that the total glycine pool measured after hydrolysis of the Bennu hot-water extract becomes available to the fluid. With \(C_{\mathrm{rock}}^{\mathrm{Gly}}=4.4\times10^{-5}\), complete release, and \(W/R=0.3\):
\[
[\mathrm{Gly}]_{f,\mathrm{RF}}^{\max}
\approx
\frac{4.4\times10^{-5}}{0.3}
=
1.47\times10^{-4}
\approx
1.5\times10^{-4}\ \mathrm{mol\,kg^{-1}}.
\]
This value is the upper hydrothermal glycine concentration used in the glycine sensitivity priors. In Fig.~2, the dashed vertical line converts this concentration to an annual hydrothermal glycine-supply endmember using the low-\(T_f\), maximum-throughput end of the Minimal prior. This follows from \(M_{\mathrm{HT},yr}=F t_{\rm yr}/[C_p(T_f-T_o)]\). Numerically, \(M_{\mathrm{HT},yr}^{\max}=1.50\times10^{12}\ \mathrm{kg\,yr^{-1}}\):
\[
S_{\mathrm{Gly,RF}}^{\max}
=
M_{\mathrm{HT},yr}^{\max}[\mathrm{Gly}]_{f,\mathrm{RF}}^{\max}
\approx
2.25\times10^8\ \mathrm{mol\,yr^{-1}}.
\]
The gray fermentable-organic glycine-equivalent proxy band in Fig.~2 converts the full proxy concentration prior through the sampled \(T_f\)-dependent hydrothermal throughput range. Its lower and upper edges are \(M_{\mathrm{HT},yr}^{\min}(10^{-8})\approx1.1\times10^3\ \mathrm{mol\,yr^{-1}}\) and \(M_{\mathrm{HT},yr}^{\max}(3\times10^{-3})\approx4.5\times10^9\ \mathrm{mol\,yr^{-1}}\), respectively.

The Standard and Serpentinization-Enhanced runs also include a wider glycine-equivalent prior:
\[
[\mathrm{Gly}]_f^{\rm eq}
\sim
\ln U\!\bigl(1\times10^{-8},\,3\times10^{-3}\bigr)\ \mathrm{mol\,kg^{-1}}.
\]
We represent fermentable organic material from these possible sources using glycine as the model substrate. We refer to the concentration assigned to this substrate as the glycine-equivalent concentration. A \(3\) mM glycine-only concentration from Bennu-like rock would require \(W/R\approx4.4\times10^{-5}/(3\times10^{-3})\approx1.5\times10^{-2}\), even for complete release. We use \(3\) mM as an exploratory upper concentration for the combined supply of glycine and other similarly fermentable organic compounds. Potential contributions include material released from the core, abiotic organic synthesis, and organic matter produced by an earlier methanogenic population \cite{SI-Steel2017,SI-Craddock2026,SI-Affholder2022}. Whether these sources could sustain this concentration remains uncertain. We calculate the fermentation products using the glycine reaction in main-text Eq. 1. The upper glycine concentration estimated from leaching remains \(1.5\times10^{-4}\ \mathrm{mol\,kg^{-1}}\).

\emph{Acetate (acetic acid) calculations.} For Bennu hot-water acetic acid \cite{SI-GlavinBennu2025},
\[
C_{\mathrm{rock}}^{\mathrm{Ac}}=1436\ \mathrm{nmol\,g^{-1}} = 1.436\times10^{-3}\ \mathrm{mol\,kg^{-1}\ rock},
\]
and \(f_{\mathrm{avail}}^{\mathrm{Ac}}\!\to\!1\), \(f_{\mathrm{leach}}^{\mathrm{Ac}}=1\). Equation~\eqref{eq:wr_map_prior} then gives
\[
W/R=10:\quad [\mathrm{Ac}]_f\approx1.436\times10^{-4}\ \mathrm{mol\,kg^{-1}}
\quad(0.144\ \mathrm{mmol\,kg^{-1}}),
\]
\[
W/R=0.3:\quad [\mathrm{Ac}]_f\approx4.79\times10^{-3}\ \mathrm{mol\,kg^{-1}}
\quad(4.79\ \mathrm{mmol\,kg^{-1}}).
\]
Bennu also contains formate, propionate, and other monocarboxylates. Using acetic acid alone \cite{SI-GlavinBennu2025}, we round the calculated upper concentration to \(5\) mmol\,kg\(^{-1}\) and extend the lower limit to \(0.05\) mmol\,kg\(^{-1}\) to explore smaller acetate supplies. This gives the prior
\[
[\mathrm{CH_3COO^-}]_f \sim \ln U\!\bigl(5\times10^{-5},\,5\times10^{-3}\bigr)\ \mathrm{mol\,kg^{-1}},
\]
i.e., \(0.05\text{--}5\ \mathrm{mmol\,kg^{-1}}\). 

\subsection*{Methods S6. Observable conversion and interval-constrained inference}
\label{sec:si_methods_s6}

For comparison with Cassini, we use annual escape-flux intervals for \ch{H2} and \ch{CH4}. Waite et al.\ report plume gas-phase volume mixing ratios of 0.4--1.4\% for \ch{H2} and 0.1--0.3\% for \ch{CH4} \cite{SI-Waite2017}. We combine these gas fractions with a water-vapor ejection range of 170--300~\si{\kilogram\per\second}, from the lower UVIS occultation estimate of Hansen et al.\ \cite{SI-Hansen2011} to the later full-mission UVIS average of Hansen et al.\ \cite{SI-Hansen2020}. Water vapor accounts for 96--99\% of the plume gas by volume \cite{SI-Waite2017}. We therefore approximate the total molar gas flow by the molar water-vapor flow when converting gas abundances to annual escape fluxes. For species \(i\), this gives
\begin{equation}
F_{i}^{\rm low/high}\approx\biggl(\frac{\dot m_{\rm H_2O}}{M_{\rm H_2O}}\biggr)\,x_i^{\rm low/high}\times t_{\rm yr},
\end{equation}
where \(x_i^{\rm low}\) and \(x_i^{\rm high}\) are the lower and upper volume mixing-ratio bounds, \(\dot m_{\rm H_2O}=170\text{--}300\ \mathrm{kg\,s^{-1}}\), \(M_{\rm H_2O}=0.01801528\ \mathrm{kg\,mol^{-1}}\), and \(t_{\rm yr}=365\times24\times3600=3.1536\times10^{7}\ \mathrm{s\,yr^{-1}}\). We write these bounds as \(\Phi_{i}^{\rm obs,low/high} \equiv F_i^{\rm low/high}\). Simulated fluxes are put in the same annual units with \(\Phi_i^{\rm sim} \equiv \Phi_i \, t_{\rm yr}\). These intervals reflect the adopted ranges of plume gas fractions and water-vapor ejection rates. Inferring dissolved gas concentrations before eruption requires additional assumptions about gas exsolution and water-vapor condensation \cite{SI-Fifer2022}.

For the model output, we use the Affholder et al.\ plume-flux definition \cite{SI-Affholder2021}:
\begin{equation}
\Phi_i \equiv \int_{S} J_c(\mathbf{x})\,C_i(\mathbf{x})\,\mathrm{d}S \quad (\mathrm{mol\,s^{-1}}),
\end{equation}
For the one-dimensional annular grid used here, this becomes
\begin{equation}
\Phi_i = \int_{0}^{R_{\max}} 2\pi r\, J_c(r)\, C_i(r)\, \mathrm{d}r \quad (\mathrm{mol\,s^{-1}}),
\end{equation}
with \(C_i\) in mol\,kg\(^{-1}\), \(J_c\) in kg\,m\(^{-2}\)s\(^{-1}\), and \(R_{\max}\) the radial cutoff.

After the ODE integration, the final \(C_i(r)\) profiles are put into this integral and multiplied by \(t_{\rm yr}\).

We compare the absolute \ch{H2} and \ch{CH4} fluxes, which also determine the ratio of methane to hydrogen. The ratio therefore adds no independent information to the classifier. Using \ch{CO2} or reaction quotients would require additional assumptions about ocean composition, carbonate speciation, and plume transport \cite{SI-Affholder2021}.

\emph{Interval-constrained inference.}

The main-text counts, the parameter subset in Fig. S1, and the joint-window fractions in Figs. S2--S3 require both gas fluxes to be consistent with one water-vapor ejection rate, with that rate and both gas fractions within their adopted bounds. This produces a region with slanted sides in log-flux space because both gas fluxes scale with the same water-ejection rate. 

\subsection*{Methods S7. ABC--RF classifier and posterior summaries}
\label{sec:si_methods_s7}

The same simulation table is used for Approximate Bayesian Computation with Random Forests (ABC--RF) \cite{SI-Pudlo2016}. Each run is passed to the classifier as
\[
s(\theta)=\left(\log_{10}\Phi_{H_2},\ \log_{10}\Phi_{CH_4}\right),
\]
and no other feature is supplied. The quantity \(\log_{10}(\Phi_{H_2}/\Phi_{CH_4})\) is computed only for plotting. The classification and regression forests used 1,500 and 1,200 trees and minimum leaf sizes of 10 and 5, respectively, with square-root feature subsampling and bootstrap sampling \cite{SI-Pedregosa2011,SI-Breiman2001}. Sample weights gave each community class equal total weight, and classifier performance was assessed by five-fold cross-validation with all simulations from the same parameter sample kept in the same fold. The classifier predicts a community class for each flux pair, and a regression forest trained on the cross-validation errors estimates the probability that this prediction is incorrect. One minus that error probability gives the estimated posterior probability of the predicted class.

For main-text Figs. 3 and 4, classifier training used simulations in which each guild could grow when introduced and the model reached equilibrium after each addition (Methods S3--S4). At least one modeled location around the vent had to meet these conditions for every guild in sequence. The four classes were H, GH, GA, and pooled GHA/GAH. For each parameter sample, a three-guild simulation received a weight of 0.5 if both introduction orders met these requirements, or 1 if only that order did. These weights were then rescaled to give each class equal total weight. For each case, we generated \(N=30{,}000\) observation samples under each uncertainty model, using fixed random seeds. We sampled the water-vapor ejection rate and both gas fractions independently within their adopted bounds, using uniform or truncated-normal distributions. For the latter, the underlying normal distributions were centered on the interval midpoints, with standard deviations equal to half the interval widths; values outside the bounds were resampled. The same sampled water-ejection rate was used to calculate both annual gas fluxes, whose \(\log_{10}\) values were supplied to the classifier. In main-text Figs. 3 and 4, panel B shows the fraction of observation samples assigned to each community class. For each fit, panel C averages the estimated posterior probabilities over the samples assigned to that class. Bars show the mean across five fits, and error bars span their minimum and maximum. For panel C, fits with no samples assigned to a class are omitted; no bar is shown if that class receives no assignments in any fit. Figure S5 shows weighted confusion matrices from five-fold cross-validation across the five fits, with all simulations from a given parameter sample kept in the same fold within each fit.

\section*{Supplementary Results}

\subsection*{Supplementary Results S1. Priors compatible with observed fluxes and sensitivity-case guide}
Figure~S1 compares the full set of parameter samples with those for which completed GA or GHA simulations satisfy both Cassini-derived flux constraints in the Standard and Serpentinization-Enhanced cases. Each parameter sample is counted once. In Figs.~S2 and S3, the right-hand bars show the fraction of simulations satisfying the \ch{H2} constraint, the \ch{CH4} constraint, or both, for each community class. These fractions use the simulations and GHA/GAH weights described in Methods~S7; the joint constraint is defined in Methods~S6.

\subsection*{Supplementary Results S2. Oxidant limitations and methanotrophy}\label{sec:oxidant_methanotrophy}
Ray et al.\ distinguish three cases for radiolytic oxidants in Enceladus's ocean: Case~I, buildup in the ocean; Case~II, consumption by aqueous reductants; and Case~III, consumption at or within the seafloor. In Case~I, separate scenarios yield ocean concentrations of up to $\sim$2.5~mmol\,kg$^{-1}$ \ch{O2} or $\sim$3.1~mmol\,kg$^{-1}$ \ch{H2O2} when both ice-shell and \ch{^{40}K} radiolysis operate. With ice-shell radiolysis alone, the corresponding values are about 1.0 and 1.6~mmol\,kg$^{-1}$, respectively. Ray et al.\ argue, however, that Cases~II--III are more consistent with Cassini constraints. In those cases, oxidants are largely scavenged by Fe/S buffering near vents, sulfate remains $\lesssim$1~mmol\,kg$^{-1}$, and steady-state \ch{O2}/\ch{H2O2} in the ocean remains very low across parameter sweeps \cite{SI-Ray2021}.

\subsubsection*{Bounds based on Ray et al. (2021) oxidant budget}
Ray et al.\ tabulate cumulative oxidant production and delivery over Enceladus's history. Their upper values are $9.4\times10^{15}$~mol \ch{O2} and $3.3\times10^{16}$~mol \ch{H2O2} from the ice shell, plus $4.2\times10^{16}$~mol \ch{O2} and $1.4\times10^{15}$~mol \ch{H2O2} produced in the ocean by \ch{^{40}K} radiolysis \cite{SI-Ray2021}. Dividing these totals by $4.5$~Gyr gives upper estimates of the average amounts of \ch{O2} and \ch{H2O2} supplied per year:
\[
\dot n_{\ch{O2}}\approx 1.14\times10^{7}\ \si{\mol\per\year},\qquad
\dot n_{\ch{H2O2}}\approx 7.6\times10^{6}\ \si{\mol\per\year}.
\]
In Cases~II--III, only a fraction of this budget reaches any single sink. Near-vent oxidants are consumed by dissolved \ch{HS^-}, dissolved \ch{Fe^{2+}}, and reduced Fe--S minerals at or within the seafloor \cite{SI-Ray2021}.

\subsubsection*{Stoichiometric constraints on methane oxidation}
Complete methane oxidation requires two moles of \ch{O2} or four moles of \ch{H2O2} per mole of \ch{CH4}:
\begin{align}
\ch{CH4 + 2 O2} &\ch{-> CO2 + 2 H2O},\label{eq:ch4_o2_sup}\\
\ch{CH4 + 4 H2O2} &\ch{-> CO2 + 6 H2O}.\label{eq:ch4_h2o2_sup}
\end{align}
Sustained sulfate-mediated anaerobic oxidation of methane implies the same net oxidant demand at system scale, because the \ch{HS^-} produced must be re-oxidized to regenerate sulfate (e.g., \ch{HS^- + 2 O2 -> SO4^{2-} + H+}). We calculate an upper bound on the average amount of methane that could be oxidized per year, assuming the entire oxidant supply is used for methane oxidation:
\[
\Phi_{\mathrm{CH_4}}^{\max}=\tfrac12\dot n_{\ch{O2}}+\tfrac14\dot n_{\ch{H2O2}}\approx 7.6\times10^{6}\ \si{\mol\per\year}.
\]
We represent sulfate-coupled anaerobic oxidation of methane with a consolidated reaction in aqueous form:
\[
\ch{CH4 + SO4^{2-} -> HCO3^- + HS^- + H2O}.
\]
At the pH range explored here (9--11), \ch{SO4^{2-}}\,$\gg$\,\ch{HSO4^-} and \ch{HS^-}\,$\gg$\,\ch{H2S}. We therefore keep the chemistry in sulfate/sulfide species form in the main text. In the state vector, the oxidant- and reductant-equivalent pools are represented as \([\mathrm{H_2SO_4}]\) and \([\mathrm{H_2S}]\).

\subsubsection*{Results and Discussion}
Cassini INMS \ch{CH4} mixing ratios (about 0.1--0.3\%), combined with the adopted water-vapor envelope of 170--300~\si{\kilogram\per\second}, imply a plume methane escape of roughly $(3.0$--$15.8)\times10^{8}$~\si{\mol\per\year} \cite{SI-Waite2017,SI-Hansen2011,SI-Hansen2020}. The upper bound on the methane oxidation rate averaged over 4.5 Gyr, \(\Phi_{\mathrm{CH_4}}^{\max}\approx 7.6\times10^{6}\)~\si{\mol\per\year}, is about 0.5--2.6\% of the methane escape flux inferred from Cassini:
\[
\frac{\Phi_{\mathrm{CH_4}}^{\max}}{3.0\times10^{8}}\approx 2.6\%\quad\text{to}\quad
\frac{\Phi_{\mathrm{CH_4}}^{\max}}{1.58\times10^{9}}\approx 0.5\%.
\]

\subsection*{Supplementary Results S3. Vent-scale habitability enablement and area-based extent}
The main text reports a vent-scale binary habitability metric for each guild defined using viability of guilds at equilibrium. A prior sample's habitability is counted as enabled when the glycine-fermenter stage creates at least one vent location that is newly viable for hydrogenotrophic methanogenesis, acetoclastic methanogenesis, or either guild. We use viability because, after the \(G\) stage, \(H\) and \(A\) can be evaluated for habitability before their later introduction in the staged assembly. We also count cases in which both guilds gain newly viable habitat somewhere in the same simulation; those gains may occur in different radial shells. Thus the vent-scale count asks whether fermentation opens new methanogen habitat anywhere in the modeled vent.

Each location in the modeled vent is simulated independently, with no diffusion between locations. For each methanogenic guild, we add up the areas where it can maintain or increase its biomass and divide by the total modeled vent area. Let \(w_r\) denote the area fraction associated with radial location \(r\), with \(\sum_r w_r = 1\). Let \(Z\in\{H, A, H\cap A\}\), where \(H\cap A\) denotes simultaneous local viability of hydrogenotrophs and acetoclastic methanogens in the same radial shell. Let \(V_r^{\mathrm{abiotic}}(Z)\) indicate that target \(Z\) is viable at shell \(r\) under the abiotic background, \(V_r^{\mathrm{afterG}}(Z)\) indicate viability at the same shell after the glycine-fermenter stage, and \(V_r^G\) indicate local viability of glycine fermenters after that stage. The indicator operator \(\mathbf{1}[\cdot]\) equals 1 when the statement in brackets is true and 0 otherwise. We then define
\[
P_{\mathrm{pre}}(Z) = \sum_r w_r \, \mathbf{1}\!\left[V_r^{\mathrm{abiotic}}(Z)\right],
\]
\[
P_{\mathrm{post}}(Z) = \sum_r w_r \, \mathbf{1}\!\left[V_r^{\mathrm{afterG}}(Z)\right],
\]
\[
P_{\mathrm{enabled}}(Z) = \sum_r w_r \, \mathbf{1}\!\left[V_r^{\mathrm{afterG}}(Z)\right]\left(1-\mathbf{1}\!\left[V_r^{\mathrm{abiotic}}(Z)\right]\right),
\]
\[
P_{\mathrm{enabled},G}(Z) = \sum_r w_r \, \mathbf{1}\!\left[V_r^G\right]\mathbf{1}\!\left[V_r^{\mathrm{afterG}}(Z)\right]\left(1-\mathbf{1}\!\left[V_r^{\mathrm{abiotic}}(Z)\right]\right).
\]
Here \(P_{\mathrm{pre}}(Z)\) is the abiotic area fraction habitable for \(Z\), and \(P_{\mathrm{post}}(Z)\) is the total area fraction habitable for \(Z\) after the glycine-fermenter stage. The term \(P_{\mathrm{enabled}}(Z)\) is the newly created post-\(G\) habitable area. The term \(P_{\mathrm{enabled},G}(Z)\) is the subset of that newly enabled area where \(G\) is also locally viable at the end of the \(G\) stage. Newly enabled habitat is viable after \(G\) and was not viable before \(G\). Net habitat change is \(P_{\mathrm{post}}(Z)-P_{\mathrm{pre}}(Z)\). Exclusion or loss corresponds to shells that are viable before \(G\) and lose viability after \(G\). The vent-scale metric records whether new habitat appears anywhere in the vent. The area fractions measure how much habitat is gained or lost and the net change.

In the Minimal case, abiotic habitable area for \(H\), \(A\), and \(H\cap A\) is zero. All post-\(G\) habitat is newly enabled. At the vent scale, fermentation enables either methanogenic guild somewhere in the vent in \(19{,}907/29{,}970=66.4\%\) of parameter samples. Hydrogenotrophs are enabled in \(35.7\%\), acetoclastic methanogens in \(66.4\%\), and both guilds gain newly viable habitat somewhere in the same simulation in \(35.7\%\). Mean newly enabled area fractions are \(0.21\%\) for \(H\), \(0.96\%\) for \(A\), and \(0.21\%\) for \(H\cap A\). The corresponding mean area fractions where \(G\) is also locally viable are \(0.10\%\), \(0.50\%\), and \(0.10\%\). On average, fermentation creates more habitable area for acetoclastic methanogens than for hydrogenotrophs.

Under Standard priors, abiotic habitable area is already nonzero before the glycine-fermenter stage with means \(P_{\mathrm{pre}}=0.004\%\) for \(H\) and \(H\cap A\), and \(1.74\%\) for \(A\). At the vent scale, fermentation enables either methanogenic guild somewhere in the vent in \(2{,}804/29{,}994\approx9.3\%\) of parameter samples. Hydrogenotrophs are enabled in \(7.3\%\), acetoclastic methanogens in \(2.6\%\), and both guilds gain newly viable habitat somewhere in the same simulation in \(0.58\%\). The added area is small on average: \(P_{\mathrm{enabled}}=0.020\%\) for \(H\) and \(H\cap A\), and \(0.003\%\) for \(A\). The corresponding mean area fractions where \(G\) is also locally viable are \(0.010\%\) for \(H\) and \(H\cap A\), and \(0.002\%\) for \(A\).

Under Serpentinization-Enhanced priors, mean abiotic habitable area is \(P_{\mathrm{pre}}=0.067\%\) for \(H\) and \(H\cap A\), and \(1.57\%\) for \(A\). At the vent scale, fermentation enables either methanogenic guild somewhere in the vent in \(2{,}023/29{,}942\approx6.8\%\) of parameter samples. Hydrogenotrophs are enabled in \(1.9\%\), acetoclastic methanogens in \(5.2\%\), and both guilds gain newly viable habitat somewhere in the same simulation in \(0.32\%\). The mean newly enabled area fractions are \(P_{\mathrm{enabled}}=0.002\%\) for \(H\) and \(H\cap A\), and \(0.007\%\) for \(A\). The corresponding mean area fractions where \(G\) is also locally viable are \(0.001\%\) for \(H\), \(0.004\%\) for \(A\), and \(0.001\%\) for \(H\cap A\). In both Standard and Serpentinization-Enhanced cases, the mean area fractions for \(H\) and \(H\cap A\) are equal, so the area suitable for both guilds is limited by hydrogenotroph viability.

\enlargethispage{\baselineskip}
\subsection*{Supplementary Results S4. Assembly-order and competitive exclusion sensitivity}
We compared GAH and GHA simulations for the same parameter samples, using only samples for which both simulations completed. The comparison included \(29{,}970\) pairs in the Minimal case, \(29{,}994\) under Standard priors, and \(29{,}942\) under Serpentinization-Enhanced priors. The final community composition, defined in Methods~S4, agreed in \(100.0\%\), \(100.0\%\), and \(99.9\%\) of pairs, respectively, rounded to one decimal place (Fig.~S6). 

In the Standard case, we compared the area viable for each methanogenic guild before and after introducing the other guild. In GHA simulations, the area fraction viable for hydrogenotrophs is already very small before acetoclastic methanogens are introduced (mean \(1.12\times10^{-4}\) of the modeled area), and falls to zero afterward. In GAH simulations, the area fraction viable for acetoclastic methanogens changes little after hydrogenotrophs are introduced (mean \(1.04\times10^{-2}\) before, \(1.02\times10^{-2}\) after). Adding glycine fermenters changes the area viable for hydrogenotrophs from zero to nonzero in \(2{,}029/29{,}994\approx6.8\%\) of parameter samples.\par These growth-area measurements are distinct from biomass persistence, assessed using the biomass threshold and equilibration requirement in Methods~S4. Under Standard priors, hydrogenotrophs met that persistence criterion alongside glycine fermenters in \(2{,}295\) of the \(29{,}994\) completed parameter samples, but in none after acetoclastic methanogens were tested in the next stage. In the reverse introduction order, no Standard sample contained a location where hydrogenotrophs passed the non-negative-growth introduction test after the stage with glycine fermenters and acetoclastic methanogens.\par Across all three cases, plume-flux differences between paired GAH and GHA simulations are small for nearly all parameter samples (Fig.~S6). Median absolute differences in \(\log_{10}\Phi_{H_2}\) and \(\log_{10}\Phi_{CH_4}\), respectively, are \((1.04\times10^{-12},7.82\times10^{-12})\) dex for Minimal, \((3.42\times10^{-13},2.55\times10^{-11})\) dex for Standard, and \((3.91\times10^{-14},1.48\times10^{-11})\) dex for Serpentinization-Enhanced priors. No \ch{H2} difference and no Standard or Serpentinization-Enhanced \ch{CH4} difference exceeds 0.001 dex. In the Minimal case, 17 of \(29{,}970\) \ch{CH4} differences exceed 0.001 dex, including four above the 0.01-dex comparison margin; the maximum is approximately 0.04 dex, equivalent to a difference of about 10\% in methane flux.

\clearpage
\begin{sifigure}
  \centering
  \safeincludegraphics[width=0.98\textwidth]{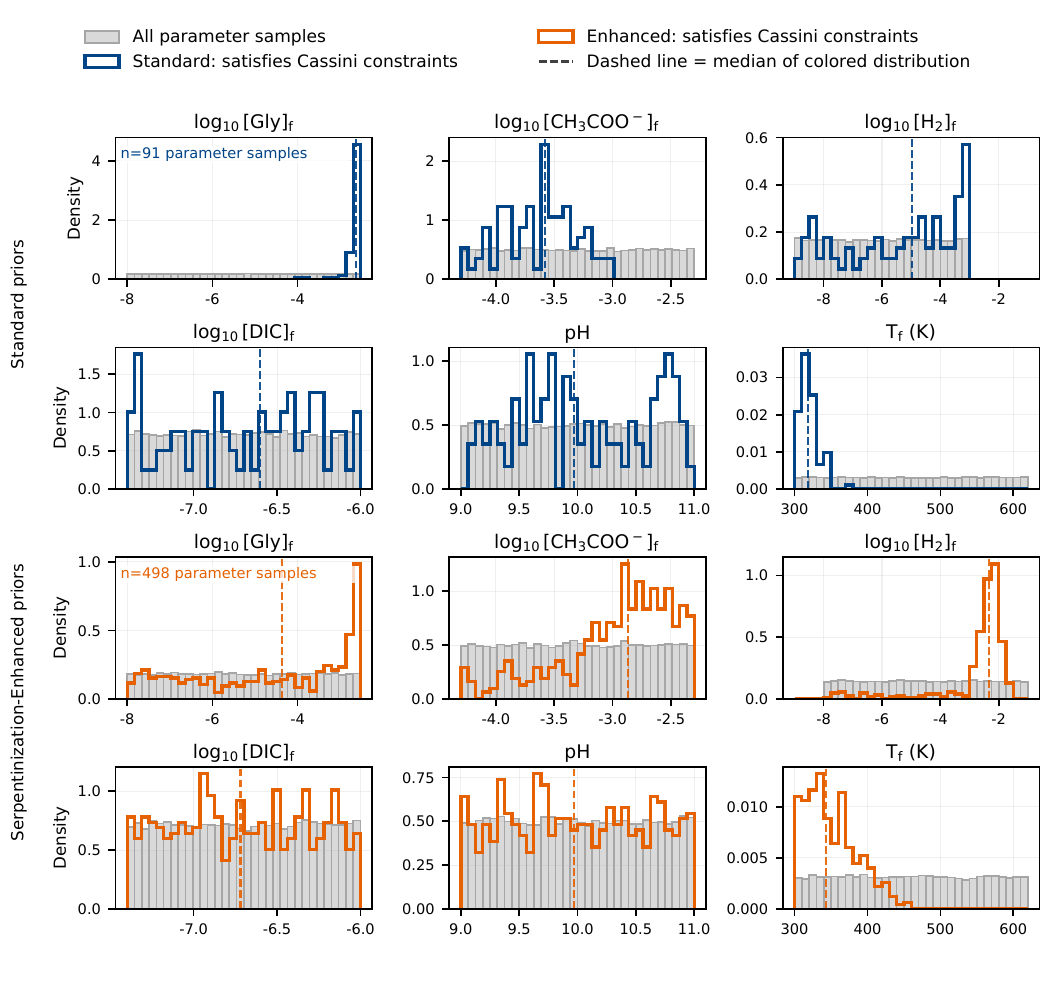}
  \caption{\textbf{Prior and Cassini-compatible distributions.}
Gray histograms show all sampled parameter values. Colored outlines show the parameter samples for which at least one completed GA or GHA simulation satisfies the joint Cassini constraint defined in Methods S6. Each parameter sample is counted once, even when both simulations satisfy the constraint. The panels show \([\mathrm{Gly}]_f\), \([\mathrm{CH_3COO^-}]_f\), \([\mathrm{H_2}]_f\), \([\mathrm{DIC}]_f\), pH, and \(T_f\). In this figure, GHA denotes simulations in which glycine fermenters are followed by hydrogenotrophic and then acetoclastic methanogens. Figure S6 compares the two orders of methanogen introduction, with glycine fermenters introduced first in both. The blue Standard samples lie toward the high-glycine end of the explored range and at low-to-moderate abiotic \ch{H2}. The orange Serpentinization-Enhanced samples extend to lower glycine because stronger abiotic \ch{H2} is allowed. The following ranges describe exported products from the qualifying GA and GHA simulations, including both results when both simulations satisfy the joint constraint. Product fluxes are reported in mol\,s\(^{-1}\); the \ch{H2}/\ch{CH4} ratio is dimensionless. Under Standard priors, the 5--95\% ranges are 1.00--15.0 for acetate, 22.7--184 for DIN, 328--\(4.69\times10^3\) for DIC, 5.29--45.6 for residual glycine, and 1.53--2.96 for the \(\Phi_{H_2}/\Phi_{CH_4}\) ratio. Under Serpentinization-Enhanced priors, the corresponding ranges are 1.13--21.5 for acetate, \(3.27\times10^{-2}\)--81.7 for DIN, 153--\(3.60\times10^3\) for DIC, \(7.53\times10^{-4}\)--25.9 for residual glycine, and 1.62--11.9 for the \(\Phi_{H_2}/\Phi_{CH_4}\) ratio.}
\end{sifigure}

\begin{sifigure}
  \centering
  \safeincludegraphics[width=0.98\textwidth]{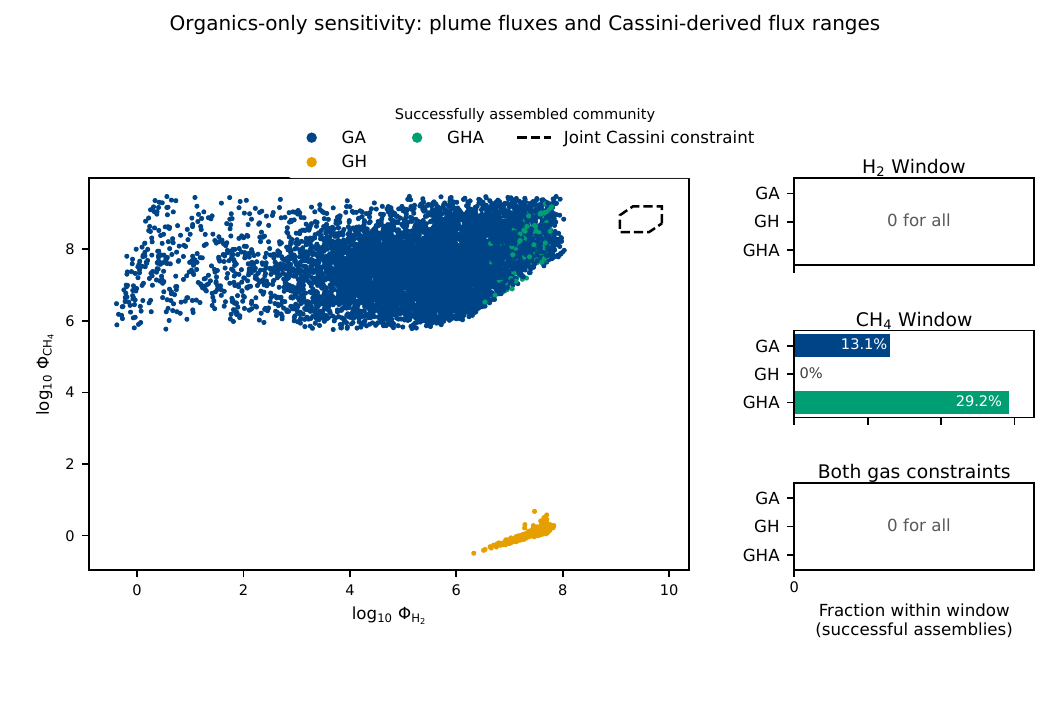}
\caption{\textbf{Organics-only sensitivity with glycine and acetate supply but no abiotic \ch{H2} or \ch{CH4}.}
The left panel shows simulated \(\log_{10}\Phi_{H_2}\)--\(\log_{10}\Phi_{CH_4}\) plume fluxes, colored by successfully assembled community. The GHA class includes both orders of methanogen introduction. The dashed outline marks the joint Cassini constraint defined in Methods S6. The right panels show, for each community class, the fractions of successfully assembled simulations satisfying the \ch{H2} constraint, the \ch{CH4} constraint, or both, calculated as described in Methods S7. In this treatment, the conservative glycine and acetate priors are retained, while abiotic \ch{H2} and abiotic \ch{CH4} are suppressed. Some GA and GHA simulations produce methane fluxes within the Cassini-derived range, but none satisfy the joint hydrogen and methane constraint.}
\end{sifigure}

\clearpage
\begin{sifigure}
  \centering
  \safeincludegraphics[width=0.98\textwidth]{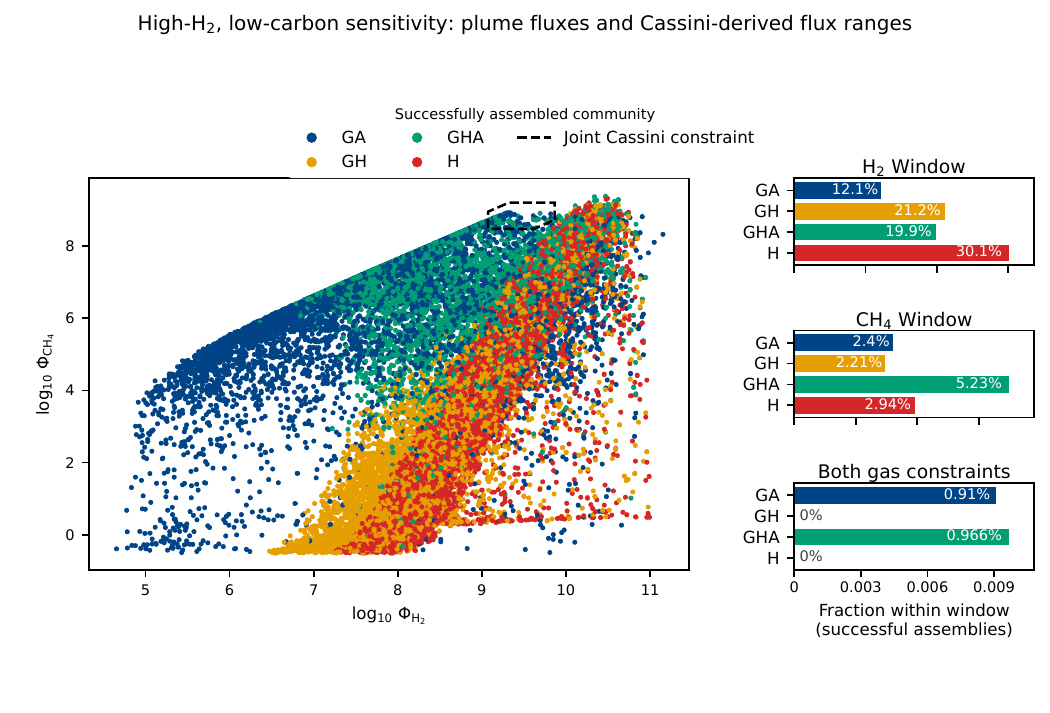}
  \caption{\textbf{High-\ch{H2}, low-carbon sensitivity without abiotic acetate or methane.}
The left panel shows simulated \(\log_{10}\Phi_{H_2}\)--\(\log_{10}\Phi_{CH_4}\) plume fluxes, colored by successfully assembled community. The GHA class includes both orders of methanogen introduction. The dashed outline marks the joint Cassini constraint defined in Methods S6. The right panels show, for each community class, the fractions of successfully assembled simulations satisfying the \ch{H2} constraint, the \ch{CH4} constraint, or both, calculated as described in Methods S7. In this variant, elevated abiotic \ch{H2} is retained while abiotic acetate and abiotic \ch{CH4} are suppressed. Some GA and GHA simulations satisfy the joint hydrogen and methane constraint; none of the GH or H simulations do.}
\end{sifigure}

\clearpage
\begin{sifigure}
  \centering
  \safeincludegraphics[width=0.98\textwidth]{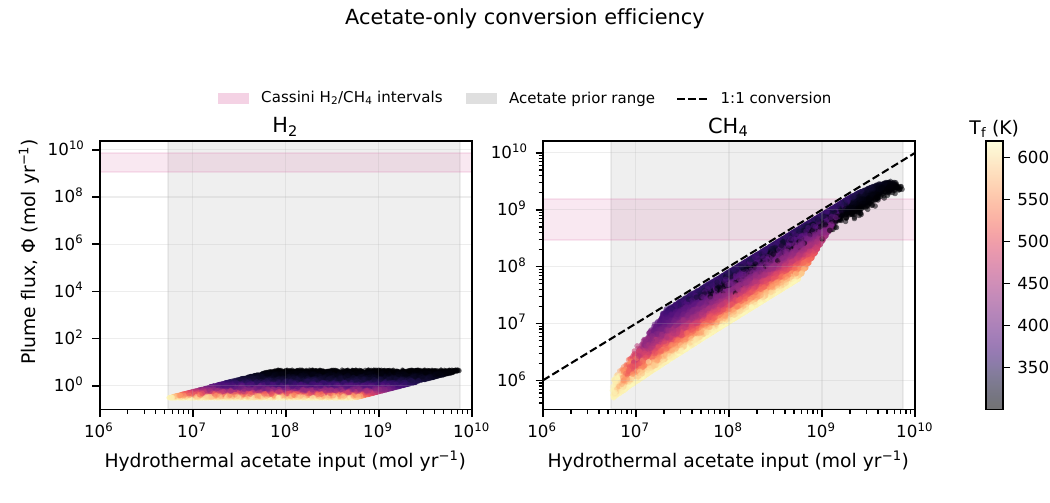}
  \caption{\textbf{Acetate-only conversion efficiency.}
Annual \ch{H2} and \ch{CH4} plume escape fluxes are plotted against the total hydrothermal acetate-supply metric \(M_{\mathrm{HT},yr}[\mathrm{CH_3COO^-}]_f\) (mol\,yr$^{-1}$) for the acetate-only treatment. Here \(M_{\mathrm{HT},yr}\) is the annual hydrothermal mass throughput defined in Methods, and \([\mathrm{CH_3COO^-}]_f\) is the hydrothermal acetate concentration. Points are colored by hydrothermal-fluid temperature \(T_f\). Pink horizontal bands mark the Cassini-derived observation intervals for each gas. The gray vertical band shows the acetate input range calculated from the acetate concentration bounds and the modeled range of hydrothermal mass throughput. The dashed line in the methane panel marks one mole of plume methane per mole of acetate supplied. The methane flux reaches the observed \ch{CH4} band over part of the acetate-supply range. \ch{H2}, however, remains near zero and far below the observed \ch{H2} band. The acetate-only treatment can reproduce the observed methane flux, while explaining both gases would require an additional hydrogen source.}
\end{sifigure}

\clearpage
\begin{sifigure}
  \centering
  \safeincludegraphics[width=0.98\textwidth]{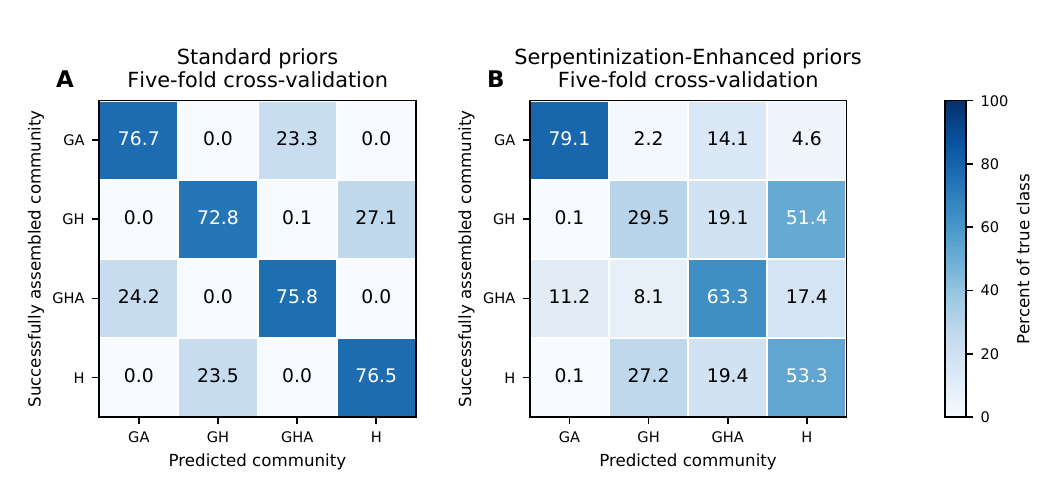}
  \caption{\textbf{ABC--RF cross-validation confusion matrices.}
Five-fold cross-validation results are shown for the random-forest classifiers used in the ABC--RF posterior summaries in main-text Figs. 3 and 4. Panel A corresponds to the Standard-prior classifier. Panel B corresponds to the Serpentinization-Enhanced-prior classifier. Rows show successfully assembled community classes; columns show the classes predicted from \((\log_{10}\Phi_{H_2},\log_{10}\Phi_{CH_4})\). Each cell gives the percentage of the row class assigned to the column class, calculated as described in Methods~S7. Diagonal cells show correct classifications; off-diagonal cells show misclassifications.}
\end{sifigure}

\clearpage
\begin{sifigure}
  \centering
  \safeincludegraphics[width=0.98\textwidth]{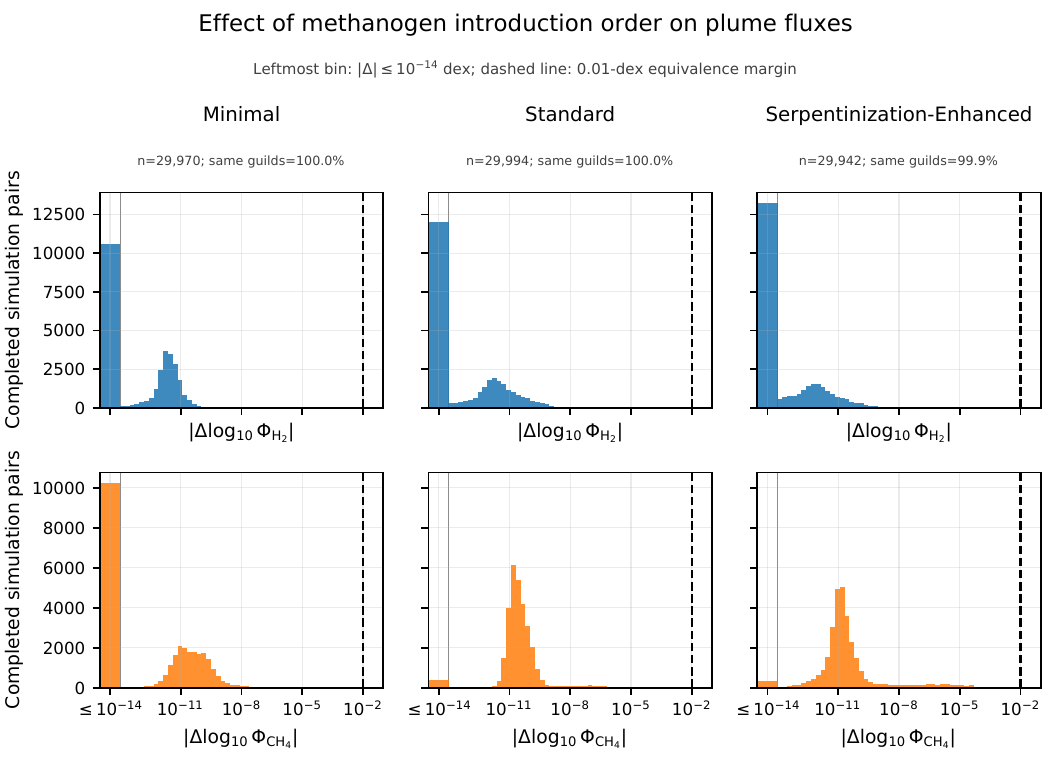}
  \caption{\textbf{Effect of methanogen introduction order on plume fluxes.}
Paired absolute differences in \(\log_{10}\Phi_{H_2}\) (top row) and \(\log_{10}\Phi_{CH_4}\) (bottom row) between GAH and GHA simulations are shown for the Minimal, Standard, and Serpentinization-Enhanced cases. Each pair uses the same parameter sample. Glycine fermenters are introduced first in both orders; acetoclastic methanogens are introduced before hydrogenotrophic methanogens in GAH and after them in GHA. Column headings give the number of completed simulation pairs and the percentage with the same guilds persisting at equilibrium (Methods S3--S4), rounded to one decimal place. The leftmost bin groups absolute differences \(\leq10^{-14}\) dex. Dashed vertical lines mark 0.01 dex, the margin used in equivalence tests of the mean signed paired difference. Four Minimal-case methane pairs exceed this margin; the largest difference is approximately 0.04 dex, equivalent to about 10\% in methane flux.}
\end{sifigure}

\clearpage
\begin{landscape}
\begingroup
\small
\setlength{\tabcolsep}{4pt}
\renewcommand{\arraystretch}{1.15}
\setlength{\LTcapwidth}{\linewidth}
\begin{longtable}{@{}L{2.8cm} C{3.4cm} C{4.4cm} C{4.4cm} L{4cm}@{}}
\caption{\textbf{Central priors.} Columns report the Minimal, Standard (with limited serpentinization), and Serpentinization-Enhanced cases. Unless stated otherwise, concentration bounds are in mol\,kg$^{-1}$; \(T_f\) is in K and pH is dimensionless. Standard gas bounds, organic concentration ranges, and fixed near-zero inputs are choices of this work, explained in Methods S5. The Serpentinization-Enhanced gas ranges follow Affholder et al. \cite{SI-Affholder2021}. Methods S5 also distinguishes measured organic inventories from their conversion to assumed fluid concentrations.}\\
\toprule
\textbf{Parameter} & \textbf{Minimal} & \textbf{Standard} & \shortstack[c]{\textbf{Serpentinization-}\\\textbf{Enhanced}} & \textbf{Ref} \\
\midrule
\endfirsthead
\multicolumn{5}{@{}l}{\textbf{Table S1 (continued)}}\\
\toprule
\textbf{Parameter} & \textbf{Minimal} & \textbf{Standard} & \shortstack[c]{\textbf{Serpentinization-}\\\textbf{Enhanced}} & \textbf{Ref} \\
\midrule
\endhead
$[\mathrm{H_2}]_f$ & Fixed $10^{-12}$ & \lnUwrap{10^{-9}}{10^{-3}} & \lnUwrap{10^{-8}}{10^{-1}} & Methods S5; \cite{SI-Affholder2021} \\
$[\mathrm{H_2}]_o$ & Fixed $10^{-12}$ & \lnUwrap{10^{-10}}{3\times10^{-7}} & \lnUwrap{10^{-8}}{10^{-6}} & Methods S5; \cite{SI-Affholder2021} \\
$[\mathrm{CH_4}]_f$ & Fixed $10^{-12}$ & \lnUwrap{10^{-12}}{10^{-6}} & \lnUwrap{10^{-8}}{10^{-4}} & Methods S5; \cite{SI-Affholder2021} \\
$[\mathrm{CH_4}]_o$ & Fixed $10^{-12}$ & \lnUwrap{10^{-12}}{10^{-7}} & \lnUwrap{10^{-8}}{10^{-6}} & Methods S5; \cite{SI-Affholder2021} \\
$[\mathrm{Gly}]_f$ & \lnUwrap{10^{-8}}{10^{-1}} & \lnUwrap{10^{-8}}{3\times10^{-3}} & \lnUwrap{10^{-8}}{3\times10^{-3}} & Methods S5 \\
$[\mathrm{Gly}]_o$ & Fixed $10^{-12}$ & \lnUwrap{10^{-10}}{10^{-6}} & \lnUwrap{10^{-10}}{10^{-6}} & Methods S5 \\
$[\mathrm{CH_3COO^-}]_f$ & Fixed $10^{-12}$ & \lnUwrap{5\times10^{-5}}{5\times10^{-3}} & \lnUwrap{5\times10^{-5}}{5\times10^{-3}} & Methods S5 \\
$[\mathrm{CH_3COO^-}]_o$ & Fixed $10^{-12}$ & \lnUwrap{10^{-8}}{10^{-5}} & \lnUwrap{10^{-8}}{10^{-5}} & Methods S5 \\
$[\mathrm{DIC}]_f$ & \lnUwrap{4\times10^{-8}}{10^{-6}} & \lnUwrap{4\times10^{-8}}{10^{-6}} & \lnUwrap{4\times10^{-8}}{10^{-6}} & \cite{SI-Affholder2021} \\
$[\mathrm{DIC}]_o$ & \lnUwrap{4\times10^{-3}}{10^{-1}} & \lnUwrap{4\times10^{-3}}{10^{-1}} & \lnUwrap{4\times10^{-3}}{10^{-1}} & \cite{SI-Affholder2021} \\
$[\mathrm{DIN}]_{o,f}$ & \lnUwrap{7.05\times10^{-9}}{2.87\times10^{-3}} & \lnUwrap{7.05\times10^{-9}}{2.87\times10^{-3}} & \lnUwrap{7.05\times10^{-9}}{2.87\times10^{-3}} & \cite{SI-Affholder2025} \\
$[\mathrm{H_2SO_4}]_{o,f}$ & Fixed $10^{-12}$ & Fixed $10^{-12}$ & Fixed $10^{-12}$ & This work \\
$[\mathrm{H_2S}]_{o,f}$ & Fixed $10^{-12}$ & Fixed $10^{-12}$ & Fixed $10^{-12}$ & This work \\
$T_f$ & \Uwrap{300}{620} & \Uwrap{300}{620} & \Uwrap{300}{620} & \cite{SI-Affholder2021} \\
pH & \Uwrap{9}{11} & \Uwrap{9}{11} & \Uwrap{9}{11} & Methods S5; \cite{SI-Affholder2025} \\
\bottomrule
\end{longtable}
\endgroup
\end{landscape}

\clearpage
\begin{landscape}
\begingroup
\small
\setlength{\tabcolsep}{4pt}
\renewcommand{\arraystretch}{1.15}
\setlength{\LTcapwidth}{\linewidth}
\begin{longtable}{@{}L{2.5cm} C{3.2cm} C{3.2cm} C{3.2cm} C{3.2cm} L{3.5cm}@{}}
\caption{\textbf{Sensitivity priors.} Columns report the high-H$_2$ glycine-only, glycine-only, acetate-only, and organics-only (glycine+acetate) treatments. Unless stated otherwise, concentration bounds are in mol\,kg$^{-1}$; \(T_f\) is in K and pH is dimensionless. The high-H$_2$ ranges follow Affholder et al. \cite{SI-Affholder2021}. Organic concentration ranges and fixed near-zero inputs are choices of this work; their rationale is given in Methods S5.}\\
\toprule
\textbf{Parameter} & \shortstack[c]{\textbf{High-H$_2$}\\\textbf{glycine-only}} & \textbf{Glycine-only} & \textbf{Acetate-only} & \shortstack[c]{\textbf{Organics-only}\\\textbf{(glycine+acetate)}} & \textbf{Ref} \\
\midrule
\endfirsthead
\multicolumn{6}{@{}l}{\textbf{Table S2 (continued)}}\\
\toprule
\textbf{Parameter} & \shortstack[c]{\textbf{High-H$_2$}\\\textbf{glycine-only}} & \textbf{Glycine-only} & \textbf{Acetate-only} & \shortstack[c]{\textbf{Organics-only}\\\textbf{(glycine+acetate)}} & \textbf{Ref} \\
\midrule
\endhead
$[\mathrm{H_2}]_f$ & \lnUwrap{10^{-8}}{10^{-1}} & Fixed $10^{-12}$ & Fixed $10^{-12}$ & Fixed $10^{-12}$ & \cite{SI-Affholder2021} \\
$[\mathrm{H_2}]_o$ & \lnUwrap{10^{-8}}{10^{-6}} & Fixed $10^{-12}$ & Fixed $10^{-12}$ & Fixed $10^{-12}$ & \cite{SI-Affholder2021} \\
$[\mathrm{CH_4}]_f$ & Fixed $10^{-12}$ & Fixed $10^{-12}$ & Fixed $10^{-12}$ & Fixed $10^{-12}$ & This work \\
$[\mathrm{CH_4}]_o$ & Fixed $10^{-12}$ & Fixed $10^{-12}$ & Fixed $10^{-12}$ & Fixed $10^{-12}$ & This work \\
$[\mathrm{Gly}]_f$ & \lnUwrap{10^{-8}}{3\times10^{-3}} & \lnUwrap{10^{-8}}{1.5\times10^{-4}} & Fixed $10^{-12}$ & \lnUwrap{10^{-8}}{1.5\times10^{-4}} & Methods S5 \\
$[\mathrm{Gly}]_o$ & \lnUwrap{10^{-10}}{10^{-6}} & \lnUwrap{10^{-10}}{10^{-6}} & Fixed $10^{-12}$ & \lnUwrap{10^{-10}}{10^{-6}} & Methods S5 \\
$[\mathrm{CH_3COO^-}]_f$ & Fixed $10^{-12}$ & Fixed $10^{-12}$ & \lnUwrap{5\times10^{-5}}{5\times10^{-3}} & \lnUwrap{5\times10^{-5}}{5\times10^{-3}} & Methods S5 \\
$[\mathrm{CH_3COO^-}]_o$ & Fixed $10^{-12}$ & Fixed $10^{-12}$ & \lnUwrap{10^{-8}}{10^{-5}} & \lnUwrap{10^{-8}}{10^{-5}} & Methods S5 \\
$[\mathrm{DIC}]_f$ & \lnUwrap{4\times10^{-8}}{10^{-6}} & \lnUwrap{4\times10^{-8}}{10^{-6}} & \lnUwrap{4\times10^{-8}}{10^{-6}} & \lnUwrap{4\times10^{-8}}{10^{-6}} & \cite{SI-Affholder2021} \\
$[\mathrm{DIC}]_o$ & \lnUwrap{4\times10^{-3}}{10^{-1}} & \lnUwrap{4\times10^{-3}}{10^{-1}} & \lnUwrap{4\times10^{-3}}{10^{-1}} & \lnUwrap{4\times10^{-3}}{10^{-1}} & \cite{SI-Affholder2021} \\
$[\mathrm{DIN}]_{o,f}$ & \lnUwrap{7.05\times10^{-9}}{2.87\times10^{-3}} & \lnUwrap{7.05\times10^{-9}}{2.87\times10^{-3}} & \lnUwrap{7.05\times10^{-9}}{2.87\times10^{-3}} & \lnUwrap{7.05\times10^{-9}}{2.87\times10^{-3}} & Methods S5; \cite{SI-Affholder2025} \\
$[\mathrm{H_2SO_4}]_{o,f}$ & Fixed $10^{-12}$ & Fixed $10^{-12}$ & Fixed $10^{-12}$ & Fixed $10^{-12}$ & This work \\
$[\mathrm{H_2S}]_{o,f}$ & Fixed $10^{-12}$ & Fixed $10^{-12}$ & Fixed $10^{-12}$ & Fixed $10^{-12}$ & This work \\
$T_f$ & \Uwrap{300}{620} & \Uwrap{300}{620} & \Uwrap{300}{620} & \Uwrap{300}{620} & \cite{SI-Affholder2021} \\
pH & \Uwrap{9}{11} & \Uwrap{9}{11} & \Uwrap{9}{11} & \Uwrap{9}{11} & Methods S5; \cite{SI-Affholder2025} \\
\bottomrule
\end{longtable}
\endgroup
\end{landscape}

\clearpage
\begin{landscape}
\begingroup
\small
\setlength{\tabcolsep}{4pt}
\renewcommand{\arraystretch}{1.1}
\setlength{\LTcapwidth}{\linewidth}
\begin{longtable}{@{}L{2.7cm} L{5.5cm} L{7.5cm} C{2.6cm} C{1.8cm}@{}}
\caption{\textbf{Biological/kinetic defaults, thermodynamic constants, physical constants, and numerical settings.}}\\
\toprule
\textbf{Parameter} & \textbf{Value} & \textbf{Description} & \textbf{Unit} & \textbf{Ref} \\
\midrule
\endfirsthead
\multicolumn{5}{@{}l}{\textbf{Table S3 (continued)}}\\
\toprule
\textbf{Parameter} & \textbf{Value} & \textbf{Description} & \textbf{Unit} & \textbf{Ref} \\
\midrule
\endhead
\multicolumn{5}{@{}l}{\emph{Biological/kinetic defaults (used in all simulations)}} \\
\midrule
$T_{\mathrm{eq}}$ & 363.15 & Enzyme activation/inactivation midpoint (90~\si{\celsius}) & \si{\kelvin} & \cite{SI-Daniel2010} \\
$\Delta G^{\ddagger}_{\mathrm{cat}}$ & \SI{72}{\kilo\joule\per\mol} & Activation Gibbs energy for the Eyring rate & \si{\kilo\joule\per\mol} & \cite{SI-Daniel2010} \\
$\Delta H_{\mathrm{eq}}$ & \SI{305}{\kilo\joule\per\mol} & Enzyme deactivation enthalpy  & \si{\kilo\joule\per\mol} & \cite{SI-Daniel2010} \\
$\tau$ & $1.73\times10^{-5}$ & Catabolic enzyme to biomass ratio & --- & \cite{SI-Affholder2021} \\
$E_{\mathrm{ana}}$ & \valstack{H: \(9.86\times10^5\)\\G: \(5.24\times10^5\)\\A: \(4.32\times10^5\)} & Metabolism-specific anabolic energy costs (H autotrophic DIC; G/A heterotrophic organics) & \si{\joule\per\mol_{C_x}} & \cite{SI-Kleerebezem2010} \\
$K_m$ & $1.0\times10^{-6}$ & Monod half-saturation constant on a mol\,kg$^{-1}$ basis & \si{\mol\per\kilogram} & (this work) \\
$d$ & $1.0\times10^{-6}$ & Baseline mortality ($\approx0.086\ \mathrm{d^{-1}}$) & \si{\per\second} & (this work) \\
$R$ & 8.314 & Universal gas constant & \si{\joule\per\mol\per\kelvin} & --- \\
$k_{B}$ & $1.380649\times10^{-23}$ & Boltzmann constant (Eyring prefactor $k_B T/h$) & \si{\joule\per\kelvin} & --- \\
$h$ & $6.62607015\times10^{-34}$ & Planck constant (Eyring prefactor $k_B T/h$) & \si{\joule\second} & --- \\
$T_S$ & 298 & Standard temperature used in thermodynamics & \si{\kelvin} & --- \\
\midrule
\multicolumn{5}{@{}l}{\emph{Base standard-state thermodynamic constants}} \\
\midrule
$\Delta G^0_S, \Delta H^0_S$ (H) & $-43.78,\,-60.54$ & Hydrogenotrophic methanogenesis base standard-state Gibbs energy and enthalpy & \si{\kilo\joule\per\mol_{eD}} & \cite{SI-Lide2005CRC} \\
$\Delta G^0_S, \Delta H^0_S$ (G) & $-5.781,\,17.4$ & Glycine fermentation base standard-state Gibbs energy and enthalpy & \si{\kilo\joule\per\mol_{eD}} & \cite{SI-Lide2005CRC} \\
$\Delta G^0_S, \Delta H^0_S$ (A) & $-55.0,\,16.2$ & Acetoclastic methanogenesis base standard-state Gibbs energy and enthalpy & \si{\kilo\joule\per\mol_{eD}} & \cite{SI-Lide2005CRC} \\
\midrule
\multicolumn{5}{@{}l}{\emph{Maintenance default}} \\
\midrule
$e_m(T)$ & \valstack{$84\,\exp\!\bigl[\tfrac{69400}{R}(1/T_S-1/T)\bigr]\cdot \tfrac{5}{432}$} & Tijhuis maintenance energy rate & \si{\joule\per\second\per\mol_{C_x}} & \cite{SI-Tijhuis1993} \\
\pagebreak
\midrule
\multicolumn{5}{@{}l}{\emph{Physical / mixing constants}} \\
\midrule
$T_o$ & 275 & Ocean temperature & \si{\kelvin} & \cite{SI-Choblet2017} \\
$F$ & $5\times10^{9}$ & Hydrothermal dissipation power & \si{\watt} & \cite{SI-Choblet2017} \\
$\varepsilon$ & 1 & Mixing-layer thickness  & \si{\meter} & \cite{SI-Affholder2021} \\
$C_p$ & 4200 & Specific heat capacity of water & \si{\joule\per\kilogram\per\kelvin} & \cite{SI-Choblet2017} \\
$\alpha$ & $3\times10^{-4}$ & Thermal expansivity of water & \si{\per\kelvin} & \cite{SI-Goodman2004} \\
$g$ & 0.12 & Enceladus surface gravity & \si{\meter\per\second\squared} & \cite{SI-Choblet2017} \\
$\rho_o$ & 1000 & Seawater density used in plume fluxes & \si{\kilogram\per\meter\cubed} & \cite{SI-Goodman2004} \\
\midrule
\multicolumn{5}{@{}l}{\textbf{Table S3 (continued). Flux-conversion constants and numerical settings.}}\\
\midrule
\multicolumn{5}{@{}l}{\emph{Flux--conversion constants (INMS $\rightarrow$ annual escape fluxes)}} \\
\midrule
$\dot{M}_{\mathrm{H_2O}}$ & 170--300 & Cassini water mass--ejection range used in flux conversion & \si{\kilogram\per\second} & \cite{SI-Hansen2011,SI-Hansen2020} \\
$M_{\mathrm{H_2O}}$ & 0.01801528 & Molar mass of water & \si{\kilogram\per\mol} & --- \\
$t_{\rm yr}$ & $3.1536\times10^{7}$ & Seconds per 365-day year used for annualization in plots & \si{\second} & (this work) \\
$M_{\mathrm{HT},yr}$ & \valstack{$M_{\mathrm{HT},yr}\equiv \dot M\,t_{\rm yr}$\\$= \dfrac{F\,t_{\rm yr}}{C_p\,(T_f - T_o)}$} & Annual hydrothermal mass throughput for the supply metrics & \si{\kilogram\per\year} & (this work) \\
\pagebreak
\midrule
\multicolumn{5}{@{}l}{\emph{Numerical domain and solver settings}} \\
\midrule
$n_r$ & 100 & Number of radial shells in the scaled annular grid & --- & (this work) \\
$x_{\mathrm{target}}$ & $10^{-5}$ & Hydrothermal mixing-ratio threshold defining \(r_{\mathrm{transition}}\) & --- & (this work) \\
$\sigma$ & 3 & Far-field radial multiplier in \(R_{\max}=\sigma r_{\mathrm{transition}}\) & --- & (this work) \\
Radial grid & scaled & 1D annular shells; \(\sim85\%\) of shells inside \(r_{\mathrm{transition}}\) & --- & (this work) \\
$R_{\max}$ & \valstack{$\sigma\,r_{\mathrm{transition}}$\\$r_{\mathrm{transition}}=c\sqrt{\tfrac{3}{2}\ln(x_0/x_{\mathrm{target}})}$} & Outer radius & \si{\meter} & (this work) \\
Solver & BDF & ODE integrator  & --- & \cite{SI-Virtanen2020} \\
rtol, atol & $10^{-7}$, $10^{-9}$ & Solver tolerances & --- & \cite{SI-Virtanen2020} \\
initial inoculum & $10^{-8}$ & Initial biomass for newly introduced metabolisms in \(B_j\) & \(\mathrm{mol\;C_x\;kg^{-1}}\) & (this work) \\
Final-community local biomass threshold & $10^{-8}(1+10^{-6})$ & Threshold on local \(B_j\) for final-community classification (Methods S4) & \(\mathrm{mol\;C_x\;kg^{-1}}\) & (this work) \\
$\Delta t_{\mathrm{check}}$ & $2\times10^{7}$ & Equilibrium-check interval between successive integration windows & \si{\second} & (this work) \\
hold time & $5\times10^{6}$ & Required quiet interval satisfying the equilibration criterion & \si{\second} & (this work) \\
min stage time & $10^{6}$ & Minimum duration before a stage can be declared equilibrated & \si{\second} & (this work) \\
$a_{\mathrm{eq}}, r_{\mathrm{eq}}$ & $10^{-9},\,10^{-7}$ & Equilibration-test absolute and relative tolerances & --- & (this work) \\
\\
\bottomrule
\end{longtable}
\endgroup
\end{landscape}

\clearpage

\renewcommand{\refname}{SI References}


\clearpage
\endgroup
\end{document}